\documentclass[11pt,oneside]{article}

\usepackage[article]{aj-article}  

\title{Galilean Anomalies and Their Effect on Hydrodynamics}

\author[a]{Akash Jain}{akash.jain@durham.ac.uk, ajainphysics@gmail.com}  

\affiliation[a]{ Centre for Particle Theory  \& Dept. of Mathematical Sciences, Durham University, UK.}

\abstract{ We extend the null background construction of \cite{Banerjee:2015uta,Banerjee:2015hra}
  to include torsion and a conserved spin current, and use it to study gauge and gravitational
  anomalies in Galilean theories coupled to torsional Newton-Cartan backgrounds. We establish that
  the relativistic anomaly inflow mechanism with an appropriately modified anomaly polynomial, can
  be used to generate these anomalies. Similar to relativistic case, we find that Galilean anomalies
  also survive only in even dimensions.  Further, these anomalies only effect the gauge and
  rotational symmetries of a Galilean theory; in particular the Milne boost symmetry remains
  non-anomalous.  We also extend the transgression machinery used in relativistic fluids to fluids
  on null backgrounds, and use it to determine how these anomalies affect the constitutive relations
  of a Galilean fluid. 

  Unrelated to Galilean fluids, we propose an analogue of the off-shell second law of thermodynamics
  for relativistic fluids introduced by \cite{Loganayagam:2011mu}, to include torsion and a conserved
  spin current in Vielbein formalism. Interestingly, we find that even in absense of spin and
  torsion the entropy currents in two formalisms are different; while the usual entropy current gets
  a contribution from gravitational anomaly, the entropy current in Vielbein formalism does not
  have any anomaly induced part.
}

\newcommand\sss{\scriptscriptstyle}
\newcommand\bsss[1]{\bar{\sss #1}}

\def\Ndot{{%
    \setbox0\hbox{$\N$}%
    \rlap{\hbox to \wd0{\hss \raisebox{4pt}{\scriptsize {$\centerdot$}}\hss}}\box0
}}
\def\Nsp{{%
    \setbox0\hbox{$\N$}%
    \rlap{\hbox to \wd0{\hss \raisebox{3pt}{\scriptsize {\textbf s}}\hss}}\box0
}}

\begin{document}

\maketitle

\section{Null Reduction and Anomalies}

The world around us, for most practical purposes, can be regarded non-relativistic. So it is
important to ask how various exotic results in relativistic theories are to be interpreted in
non-relativistic limit. Taking this limit however turns out to be a non-trivial task; except in few
special cases, non-relativistic limit is either not well defined or is not unique\footnote{For
  example, Maxwell's electromagnetism is known to have more than one non-relativistic limits
  \cite{2013EJPh...34..859M}.}, which forces the analysis to resort on approximate methods. It is
generally accepted that non-relativistic theories can be very well approximated by Galilean
theories. So rather than taking a limit of relativistic theories, one can take a more axiomatic
approach of defining the Galilean theories in their own right -- as has been historically done --
and say something useful about the non-relativistic theories. About a decade after the inception on
general relativity it was realized that Galilean spacetimes can also be packaged into a nice
covariant language -- Newton-Cartan geometries \cite{Cartan:1924yea,Cartan:1923zea}. Since then
there has been a huge amount of development in understanding how Galilean theories couple to
Newton-Cartan backgrounds
\cite{Dombrowski1964,Kuenzle:1972zw,Duval:1976ht,Kuchar:1980tw,Duval:1983pb,Duval:1984cj,2006AnPhy.321..197S,2009JPhA...42T5206D,Christensen:2013rfa,Son:2013rqa,Jensen:2014aia,Jensen:2014ama,Jensen:2014wha,Geracie:2015xfa,2015PhRvD..91d5030G}\footnote{It
  is far from the reach of a mortal being to compile an exhaustive list of work on non-relativistic
  physics; please refer to mentioned works and references therein.}. We recommend looking at \S 2.1
of \cite{Jensen:2014aia} for a short and self contained review of Newton-Cartan geometries, which
will be extensively used throughout this work. Refer
\cite{Banerjee:2014pya,Banerjee:2014nja,Brauner:2014jaa,Hartong:2014pma} for some more recent work
on Galilean physics which will not be touched upon here.

There is also a relatively recent way to approach non-relativistic physics -- \emph{null reduction}
\cite{Maldacena:2008wh,Adams:2008wt,Herzog:2008wg}. It is known for a long time that Galilean group
can be embedded into one dimensional higher Poincar\'e group. Correspondingly, one can constrain the
Poincar\'e algebra in a certain way, and reduce it to a Galilean algebra. To be more precise,
consider generators of 5 dimensional Poincar\'e algebra written in null coordinates\footnote{We
  define the transformation to null coordinates as $x^\pm = \frac{1}{\sqrt{2}} (x^0 \pm x^4)$.}
($\sA,\sB = -,+,1,2,3$),
\begin{equation}
	\text{Spacetime Translations: } P_\sA, \qquad
	\text{Lorentz Transformations: } M_{\sA\sB}.
\end{equation}
A subset of these -- generators which commute with null momenta $P_-$ ($a,b=1,2,3$),
\begin{equation} 
  P_-, \quad P_+, \quad P_a, \quad M_{a+}, \quad M_{ab}, 
\end{equation}
form a Galilean algebra, with $P_-$ acting as a new Casimir. $P_-$ can be interpreted as continuity
operator (with mass as its conserved charge), $P_a$ as translations, $P_+$ as time translation,
$M_{a+}$ as Galilean boosts and finally $M_{ab}$ as rotations (look at \cite{Rangamani:2009zz} for
an extensive review)\footnote{To be more precise, what we call Galilean group is generally known
  as the \emph{Bargmann group} which is the \emph{central extension of Galilean group} with
  central generator $P_-$. Galilean group sits inside as a special case with $P_- = 0$.}. It has far reaching implications -- one has an entire new way to take
`non-relativistic limit'. Rather than starting from a 4 dimensional relativistic theory and taking
`$c$' to infinity to get a non-relativistic theory, one can start with a 5 dimensional relativistic
theory and reduce it over a light cone (introduce a null Killing vector) to get a Galilean theory,
which is as good as a non-relativistic theory. This idea (and its generalizations to higher and
lower dimensions) have been used readily in the literature to reproduce known results and to get new
insights into non-relativistic physics. Probably most important of these was to reproduce
(torsional) Newton-Cartan geometries starting from a \emph{Bargmann structure} (relativistic
manifold carrying a covariantly constant null Killing vector) in one higher dimension
\cite{Duval:1984cj,Duval:1990hj,Julia:1994bs,Christensen:2013rfa}. Also authors in
\cite{Rangamani:2008gi} and many following them (e.g. \cite{Brattan:2010bw,Banerjee:2014mka})
established that reducing a relativistic fluid on light cone indeed gives the expected constitutive
relations of a Galilean fluid, discussed e.g. in \cite{landau1959fluid}.

In \cite{Banerjee:2014mka} authors realized that although this mechanism gives `a' Galilean fluid,
it is not the most generic one. Especially thermodynamics which the reduced fluid follows is in some
sense more restrictive than the most generic Galilean theories\footnote{See eqn. IV.121 of
  \cite{Banerjee:2014mka} and footnote (7) of \cite{Banerjee:2015hra} for more details on this
  issue.}. Also parity-violating sector of the reduced fluid is highly restrictive and survives only
in a very special case of `incompressible fluids kept in a constant magnetic field'. Same authors in
\cite{Banerjee:2015uta} provided a resolution to this issue, which however is little different from
the usual spirit of null reduction. Rather than doing null reduction of a relativistic fluid,
authors suggested to construct a theory of fluids coupled to Bargmann structures from scratch,
henceforth referred as \emph{Bargmann fluid} or \emph{null fluid}\footnote{Why `null'
  fluid? A fluid is generally called `null' if the corresponding fluid velocity is a null
  vector. Unlike usual relativistic fluids, one can show that on a Bargmann structure (with null
  Killing vector $V^\sM$), a null fluid ($u^\sM u_\sM = 0$) and a unit normalized fluid
  ($w^\sM w_\sM = -1$) are related by merely a field redefinition:
  $u^\sM = w^\sM + \frac{1}{2 w^\sN V_\sN} V^\sM$. \cite{Banerjee:2015uta} found that writing a
  Bargmann fluid in terms of `null fluid velocity' is more natural from the point of view of a
  Galilean fluid.}. In the process it was realized that there are certain aspects of null fluids
which arise just by the introduction of null isometry and have no analogue in usual relativistic
fluids. Upon null reduction\footnote{Since the theory already has a null Killing field, null
  reduction is defined as choosing a foliation transverse to the Killing field and compactifying the
  null direction. As we shall discuss in \cref{null_redn}, doing this requires introducing a
  Galilean frame of reference, or in other words a preferred notion of time.}, this null fluid gives
rise to the most generic Galilean fluid. In a sense null fluids can be seen as a particular
embedding of Galilean fluid into a spacetime of one higher dimension. This approach is more in lines
with the axiomatic approach to study Galilean theories, but has the benefit that we have all the
well-developed machinery of relativistic physics at our disposal.

The aim of this paper is to address a similar issue, but in a different setting --
\emph{anomalies}. Gauge and gravitational anomalies for a non-relativistic quantum field theory
(Lifshitz fermions) was discussed in \cite{Bakas:2011nq} using path integral
methods. \cite{Jensen:2014hqa} on the other hand took the conventional null reduction approach to
this problem, where author started with an anomalous relativistic theory and figured out its fate
upon reduction. There is however an issue with this approach -- relativistic
anomalies\footnote{Author in \cite{Jensen:2014hqa} considered both gauge/gravitational as well as
  Weyl anomalies; however in this work we will only be concerned about the former.} are known to
exist only in even dimensions, hence this approach will essentially give anomalies \emph{only in odd
  dimensional} Galilean theories. This is slightly unpleasant, because if one is to look at Galilean
theories as a makeshift for non-relativistic theories which in turn are `low velocity' limit of
relativistic theories in same number of dimensions, one would expect them to be anomalous \emph{only
  in even dimensions}. Half of this problem can be solved by noting that all the anomalies found by
\cite{Jensen:2014hqa} crucially depend on the components of higher dimensional gauge field and
affine connection along the Killing direction ($ A_{\sss\sim}$, $\G^\sM_{\ \sim\sN}$ where $A_\sM$
is the gauge field, $\G^\sR_{\ \sM\sN}$ is the affine connection, and the null Killing vector is
chosen to be $\dow_{\sss\sim}$). It was noted by \cite{Banerjee:2014mka} that these components act
as sources in the mass conservation Ward identity (look at discussion around
\cref{mass.sources}). Since we do not know of any such mass sources appearing in nature, it would be
better to switch these off (one can check that these mass sources $A_{\sss\sim}$,
$\G^\sM_{\ \sim\sN}$ are well defined gauge covariant tensors). Doing so will eliminate all the
anomalies in odd dimensional Galilean theories. We call the Bargmann structures with these mass
sources set to zero as \emph{compatible Bargmann structures} or \emph{null backgrounds}
following \cite{Banerjee:2015hra}. The other half of
the problem is however more challenging -- we need to find a consistent mechanism to introduce
anomalies in theories coupled to odd dimensional null backgrounds.

The basic idea to do this was illustrated in \cite{Banerjee:2015hra} using abelian gauge
anomalies. To motivate this lets consider the simplest case of 4 dimensional flat relativistic
theory with a $\rmU(1)$ anomaly. Conservation of corresponding (covariant) current $ J^\mu$ is given as,
\bee{
	\dow_\mu  J^\mu = \frac{3}{4} C^{(4)}  \e^{\mu\nu\r\s} F_{\mu\nu} F_{\r\s},
}
where $ F_{\mu\nu}$ is the field strength tensor and $C^{(4)}$ is the anomaly constant. Upon taking a non-relativistic limit, one would qualitatively expect the conservation law to look like (for appropriately large or small $C^{(4)}$),
\bee{
	\dow_t q + \dow_i j^i_q = 3 C^{(4)}  \ve^{ijk} F_{0i} F_{jk} = - 6 C^{(4)}  e_i b^i,
}
where $e,b$ are the electric and magnetic fields respectively. This effect can be reproduced after null reduction of a 5 dimensional conservation law,
\bee{
	\dow_\sM  J^\sM = \frac{3}{4} C^{(4)} \e^{\sM\sN\sR\sS\sT}\bar V_\sM F_{\sN\sR} F_{\sS\sT},
}
where $\bar V^\sM$ is an arbitrary null vector with $\bar V_{\sss\sim} = -1$. Note that $ F_{{\sss\sim}\sM} = \dow_{\sss\sim}  A_\sM - \dow_\sM  A_{\sss\sim} = 0$ when $ A_{\sss\sim} = 0$. Since one index on $\e$ must be `$\scriptstyle\sim$', that responsibility lands on $\bar V_\sM$ implying that the mentioned expression doesn't depend on what $\bar V^\sM$ is chosen (these statements will be made more rigorous in \cref{null_inflow}). It was observed by \cite{Banerjee:2015hra} that this anomaly can indeed be generated by anomaly inflow mechanism exactly in the same way as it works for usual relativistic anomalies, but with a tweaked anomaly polynomial. Authors there were interested in abelian anomalies and how they affect the hydrodynamics at the level of constitutive relations. This work will generalize these arguments to non-abelian and gravitational anomalies, and will give a more rigorous and transparent mechanism to compute their contribution to Galilean hydrodynamics using transgression machinery of relativistic fluids \cite{Jensen:2013kka}.

However unlike \cite{Banerjee:2015hra} we would need to introduce torsion in the game for a clearer
analysis of gravitational sector. In Newton-Cartan geometries it is known
(cref. \cite{Jensen:2014aia}) that torsionlessness imposes a dynamic constraint $\df \bm n = 0$ on
the time metric $\bm n = n_\mu \df x^\mu$. It has been noted in
\cite{Christensen:2013rfa,2015PhRvD..91d5030G,Gromov:2014vla} that lifting this constraint off-shell
is necessary to study energy transport in Galilean theories. Similar issue also showed up in the
context of Galilean hydrodynamics discussed in \cite{Banerjee:2015uta} where authors noted that on
torsionless Galilean backgrounds second law of thermodynamics fails to capture all the constraints
obeyed by transport coefficients of a Galilean fluid. Since we would be interested in off-shell
physics to understand anomalies, imposing torsionlessness would only make matters less
clear. Nevertheless, on cost of some added technicality, it would allow us to explore null reduction
for theories with non-zero spin current, which as far as we can tell hasn't been
attempted\footnote{Some authors including \cite{Christensen:2013rfa} have considered null reduction
  in presence of torsion, but have not included a spin connection as an independent background
  source.}. \cite{Geracie:2015xfa} has considered the most generic Galilean theories on torsional
Newton-Cartan background (without a conserved spin current), which follows very nicely via null
reduction. Notably authors in \cite{Geracie:2015xfa} presented their results in a
`frame-independent' manner using an `extended space representation' of the Galilean group; we show
in \cref{Geracie:2015xfa} that this representation is nothing but the theory on a null background
seen prior to null reduction.

It is worth noting here that the essence of null reduction, usual or axiomatic, lies in the fact
that the sophisticated machinery of relativistic theories can be used to say something useful about
non-relativistic theories. This method however has its limitations; one needs to be acquainted with
the relativistic side of the story to appreciate the construction. Although we review whatever is
required for this work, readers might find it helpful to consult the relativistic results first, or
from time to time during the reading. The respective relativistic references will be mentioned on
the go.

Unrelated to Galilean fluids, we also make some observations regarding entropy current for a
relativistic fluid. Recently an off-shell generalization of the second law of thermodynamics was
considered by \cite{Loganayagam:2011mu} in context of torsionless relativistic hydrodynamics. The
authors in \cite{Haehl:2014zda,Haehl:2015pja} also proposed a new abelian $\rmU(1)_\rmT$ symmetry in
hydrodynamics associated with this off-shell statement, with entropy as its conserved charge. We
propose a natural generalization of this off-shell statement of second law in Vielbein formalism, in
presence of torsion and a conserved spin current. More interestingly, even in absence of torsion we
find that the entropy current defined by off-shell second law in Vielbein formalism, is different
from what defined in usual formalism (we call the latter as Belinfante entropy current). Vielbein
entropy current does not have any anomaly induced part, while the Belinfante entropy current has
been shown to get contributions from gravitational anomaly \cite{Haehl:2015pja}. A similar
distinction between two formalisms has been known for energy-momentum tensor as well: while the
Vielbein formalism deals with an asymmetric canonical EM tensor (which is Noether current of
translations), the usual formalism deals with a symmetric Belinfante EM tensor (which couples to the
metric in general relativity) (see \cref{torsionless_EinCar_foot} for related comments). Motivated
from this, and the fact that Vielbein entropy current does not get contributions from anomaly, we
guess that it should be in some sense more naturally related to the fundamental $\rmU(1)_\rmT$
symmetry of \cite{Haehl:2014zda,Haehl:2015pja}. In the passing we would also like to note that the
two entropy currents are found to differ only off-shell, and boil down to the same on imposing
equations of motion. Further, for a spinless fluid the difference only survives in anomalous sector,
and is precisely what accounts for the Vielbein entropy current being independent of
anomalies. Interested readers can jump directly to \cref{belinEC}.

This work is broadly categorized in 5 sections. The remaining of introduction contains a summary of
our main results in \cref{results}. \cref{torsional_null} starts off by extending null background
construction of \cite{Banerjee:2015hra} to include torsion, which is further used to derive Ward
identities of a Galilean theory with non-trivial spin current in \cref{null_redn}. A review of the
relativistic anomaly inflow mechanism has been provided in \cref{anomaly_inflow}, which we modify in
\cref{null_inflow} to account for anomalies in null/Galilean backgrounds and derive corresponding
anomalous Ward identities. Later in \cref{hydro} we discuss how these anomalies affect the
constitutive relations of null/Galilean hydrodynamics. Keeping in mind the technicality of this
work, a detailed walkthrough example for the simplest case of 3 dimensional null theories (2
dimensional Galilean theories) has been given in \cref{walkthrough1}, results of which are
generalized to arbitrary higher dimensions in \cref{example_gen}. In \cref{non_cov} we present some
of our results in conventional non-covariant basis for the benefit of readers not acquainted with
Newton-Cartan language. \Cref{Geracie:2015xfa} is devoted to a comparison of null backgrounds to the
extended space representation of \cite{Geracie:2015xfa}.  In \cref{forms} we give some notations and
conventions for differential forms used throughout this work. Finally in \cref{belinEC} we comment
on the entropy current in relativistic hydrodynamics in Vielbein formalism.

\subsection{Overview and Results} \label{results}

Skipping all the technicalities we start directly with the results, keeping in mind that these
results have been obtained by null reduction of anomalies on null backgrounds. In the following we
denote indices on Newton-Cartan (NC) manifold $\cM_{(d+1)}^{\text{NC}}$ by $\mu\nu\ldots$, and on a
flat spatial manifold by $\bbR^{(d)}$ by $a,b\ldots$. NC structure is defined by a time-metric
$n_\mu$, a degenerate Vielbein $e_a^{\ \mu}$, and a flat metric $\d_{ab}$. Further we define a NC
frame velocity $v^\mu$, and using it an `inverse' Vielbein $e^a_{\ \mu}$ by
$v^\mu n_\nu + e_{a}^{\ \mu} e^a_{\ \nu} = \d^\mu_{\ \nu}$ and
$e^a_{\ \mu} e_{b}^{\ \mu} = \d^a_{\ b}$. Indices on $\cM_{(d+1)}^{\text{NC}}$ cannot be
raised/lowered, while on $\bbR^{(d)}$ can be raised/lowered by
$\d^{ab},\d_{ab}$. $\cM_{(d+1)}^{\text{NC}}$ indices can be projected down to $\bbR^{(d)}$ using
$e^a_{\ \mu}, e_{a}^{\ \mu}$. NC manifold is also equipped with a connection $\G^\l_{\ \mu\nu}$, a spin
connection $C^a_{\ \mu b}$, a non-abelian gauge field $A_\mu$ and a covariant derivative $\Ndot_\mu$
associated with all of these. Differential forms are denoted by bold symbols.

Similar to the relativistic case, we find that (gauge and gravitational) anomalies on an even
dimensional NC background $\cM^{\text{NC}}_{(2n)}$, are governed by a $(2n+2)$ dimensional anomaly
polynomial $\bm\fp^{(2n+2)}$. However here the anomaly polynomial is written in terms of Chern
classes of gauge field strength
$\bF = \df \bA + \bA \wedge \bA = \half F_{\mu\nu}\df x^\mu \wedge \df x^\nu$ and Pontryagin classes
of NC spatial curvature
$\bR^a_{\ b} = \df \bC^a_{\ b} + \bC^a_{\ c} \wedge \bC^c_{\ b} = \half
\tensor{R}{_{\mu\nu}^{a}_{b}} \df x^\mu \wedge \df x^\nu$.
The odd dimensional Galilean theories on the other hand are non-anomalous (in absence of any extra
mass source). In presence of anomalies, the conservation laws of the theory are given as,
\begin{align}
  \text{Mass Cons. (Continuity):}&\qquad \underline{\Ndot}_\mu \r^\mu = 0, \nn\\
  \text{Energy Cons. (Time Translation):}&\qquad \underline{\Ndot}_\mu \e^{\mu} = [\text{power}] - p^{\mu a} c_{\mu a}, \nn\\
  \text{Momentum Cons. (Translations):}&\qquad \underline{\Ndot}_\mu p^{\mu}_{\ a} = [\text{force}]_a
                                         - \r^{\mu} c_{\mu a}, \nn\\
  \text{Temporal Spin Cons. (Galilean Boosts):}&\qquad \underline{\Ndot}_\mu \t^{\mu a} 
                                                 = 
                                                 \half \lb \r^a - p^a\rb, \nn\\
  \text{Spatial Spin Cons. (Rotations):}&\qquad  \underline{\Ndot}_\mu \s^{\mu ab} = p^{[ba]}
                                          + 2 \t^{\mu[a} c_{\mu}^{\ b]}
                                          + {\color{red} \upsigma^{\perp ab}_{\rmH }}, \nn\\
  \text{Charge Cons. (Gauge Transformations):}&\qquad  \underline{\Ndot}_\mu j^\mu = {\color{red} \mathrm j_{\rmH }^\perp},
\end{align}
where $\underline{\Ndot}_\mu = \Ndot_\mu + v^\nu H_{\nu\mu} - e_a^{\ \nu}\rmT^a_{\ \nu\mu}$.  Here
$H_{\mu\nu}$ is the temporal torsion and $\rmT^a_{\ \mu\nu}$ is the spatial torsion. Along with the
conservation laws, the associated symmetries and conserved quantities have been specified above. We
see that mass is exactly conserved. Energy/momentum is sourced by power/force densities (expressions
can be found in \cref{null_redn}) and pseudo-power/force due to spacetime dependence of the frame
velocity $c_{\mu}^{\ a} = e^a_{\ \nu}\Ndot_{\mu}v^\nu$. Temporal spin is sourced by difference in
spatial mass current and momentum density; for spinless theories it implies equality of the two.
Barring anomalies, spatial spin is sourced by antisymmetric part of momentum density (causing
torque) and pseudo-torque, while charge is exactly conserved. In addition to these, spatial spin and
charge are also sourced by gravitational $\upsigma_{\rmH }^{\perp ab}$ and gauge $\mathrm j_{\rmH }^\perp$
anomalies respectively. These anomaly sources can be determined from the anomaly polynomial
$\bm\fp^{(2n+2)}$ as,
\begin{equation}\label{NC_Hall2_preview}
  \upsigma_{\rmH }^{\perp ab} = - \ast_\uparrow \lB \frac{\dow \bm\fp^{(2n+2)}}{\dow \bR_{ba}} \rB, \qquad
  \mathrm j_{\rmH }^\perp = - \ast_\uparrow \lB \frac{\dow \bm\fp^{(2n+2)}}{\dow \bF} \rB.
\end{equation}
In the study of Galilean hydrodynamics, we can construct the sector of constitutive
relations completely determined by mentioned anomalies. For doing this, we first need to define the
hydrodynamic shadow gauge field, $\bm{{\hat A}} = \bA - \mu \bm n$ and spin connection
$\bm{{\hat C}}{}^a_{\ b} = \bC^a_{\ b} - [\mu_{\s}]^a_{\ b} \bm n$, where $\mu$ is the gauge chemical
potential and $[\mu_{\s}]^a_{\ b}$ is the spatial spin chemical potential. We call the corresponding
field strengths $\bm{{\hat F}}$ and $\bm{{\hat R}}{}^a_{\ b}$, and the anomaly polynomial made out of
these to be $\bm{{\hat\fp}}{}^{(2n+2)}$. Using these we define the transgression form,
$\dsp \bcV_{\bm\fp}^{(2n+1)} = - \frac{\bm n}{\bH}\wedge \lb \bm\fp^{(2n+2)} -
\bm{{\hat\fp}}{}^{(2n+2)} \rb $,
where $\bH = - \df \bm n$. It can be used to generate the anomalous sector of constitutive
relations; only non-zero contributions are given as, 
\begin{equation}
  \nn (\e^\mu)_\rmA = *_\uparrow \lB \frac{\dow \bcV^{(2n+1)}_{\bm\fp}}{\dow \bH}\rB^\mu, \qquad 
  (\s^{\mu ab})_\rmA = *_\uparrow \lB \frac{\dow \bcV^{(2n+1)}_{\bm\fp}}{\dow \bR_{ba}}\rB^\mu, \qquad 
  (j^\mu)_\rmA = *_\uparrow \lB \frac{\dow \bcV^{(2n+1)}_{\bm\fp}}{\dow \bF} \rB^\mu.  
\end{equation}
We leave it for the readers to convince
themselves that these formulas are well defined. These constitutive relations follow the second law of
thermodynamics and off-shell adiabaticity with a trivially zero entropy current. We would like to
caution the reader that these are merely the contribution from anomalies to the constitutive
relations, there will be further contributions which are independent of anomalies and have not been
discussed here.

Explicit examples of the above results in case of $\rmU(1)$ and gravitational anomalies, for $2$
dimensions and a generalization to $2n$ dimensions has been given in \cref{examples}. But probably
the most important take home message of this work is that one can perform a consistent analysis of
gauge and gravitational anomalies for Galilean theories using guidelines laid out by relativistic
construction. This should be taken as a yet another point in the favor of, or rather an
advertisement for, the axiomatic approach to null reduction -- null backgrounds
\cite{Banerjee:2015uta}.

\section{Galilean Theories with Spin and Torsion} \label{torsional_null}

The aim of this section is to extend the null background construction of
\cite{Banerjee:2015uta,Banerjee:2015hra} to torsional backgrounds, and derive non-anomalous Ward
identities for a Galilean theory with non-zero spin current. We will later introduce anomalies in
\cref{anomaly_inflow}. The construction is mainly based on the work of \cite{Julia:1994bs,Christensen:2013rfa} on
torsional null reductions, with certain modifications. We will be working in Vielbein formalism,
which is most natural choice for a spin system. Hence the language and expressions will be slightly
different from what seen in the earlier work on null backgrounds \cite{Banerjee:2015hra} where
authors focus on torsionless and spinless case.

\subsection{Einstein-Cartan Backgrounds}

We start with a short review of Einstein-Cartan backgrounds, mostly to setup notation for our later
discussion on torsional null backgrounds. A more comprehensive introduction to this formalism can be
found in e.g. \cite{Watanabe:2004nt}. Consider a manifold $\cM_{(d+2)}$ theories on which are
invariant under diffeomorphisms and possibly non-abelian gauge group $\cG$. We denote the
infinitesimal diffeomorphism and gauge variation parameters by,
\begin{equation}
  \p_{\xi} = \lbr \xi = \xi^{\sM}\dow_{\sM}, \L_{(\xi)} \rbr 
  \in \rmT\cM_{(d+2)} \times \fg.
\end{equation}
We have denoted tangent bundle of $\cM_{(d+2)}$ as $\rmT\cM_{(d+2)}$, and Lie group corresponding to
$\cG$ as $\fg$. Indices on $\cM_{(d+2)}$ are denoted by $\sM,\sN,\sR,\sS\ldots$. $\cM_{(d+2)}$
is endowed with a metric $\df s^2 = \rmG_{\sss MN}\df x^{\sss M} \df x^{\sss N}$, a $\fg$ valued
gauge field $\bA = A_{\sss M} \df x^{\sss M}$ and a metric compatible affine connection
$\G^{\sss R}_{\sss \ MS}$ which is not necessarily symmetric in its last two indices.  In the case
of torsional geometries it is more natural to shift to Vielbein formalism, which we describe in the
following. The condition of local flatness of a manifold allows us to define a map between
$\rmT\cM_{(d+2)}$ and (pseudo-Riemannian) flat space $\bbR^{(d+1,1)}$, realized in terms of a
Vielbein $\rmE^{\sss A}_{\sss \ M}$ and its inverse $\rmE_{\sss A}^{\sss \ M}$, restricted by,
\begin{equation}\label{Viel_rest}
  \rmG_{\sss MN} = \rmE^{\sss A}_{\sss \ M} \rmE^{\sss B}_{\sss \ N} \eta_{\sss
    AB}, \qquad \rmG^{\sss MN} = \rmE_{\sss A}^{\sss \ M} \rmE_{\sss B}^{\sss \ N} \eta^{\sss AB}, 
\end{equation}
where $\eta_{\sss AB}$ is the flat metric, and ${\sA,\sB,\sC,\sD}\ldots$ denote indices on
$\bbR^{(d+1,1)}$. Indices on $\cM_{(d+2)}$ can be raised and lowered by $\rmG_{\sss MN}$, and on
$\bbR^{(d+1,1)}$ by $\eta_{\sss AB}$.  Indices on $\cM_{(d+2)}$ and $\bbR^{(d+1,1)}$ can also be
interchanged using the $\rmE^{\sss A}_{\sss \ M}$.  Vielbein has $(d+2)^2$ components out of which
$\half (d+2)(d+3)$ are taken away by \cref{Viel_rest}. Remaining $\half (d+1)(d+2)$ components can
be fixed by introducing an additional $\rmS\rmO(d+1,1)$ symmetry in Vielbein
$\rmE^{\sss A}_{\ \sss M} \sim O^{\sss A}_{\sss \ B} \rmE^{\sss B}_{\sss \ M}$. Hence
$\rmE^{\sss A}_{\sss \ M}$ modded by diffeomorphisms and $\rmS\rmO(d+1,1)$ has same physical information
as $\rmG_{\sss MN}$ modded with only diffeomorphisms. We also define a spin connection for fields
living in $\bbR^{(d+1,1)}$,
\begin{equation}
  \bC^{\sss A}_{\sss \ B} = C^{\sss A}_{\sss\ MB} \df x^{\sss M} =
  \rmE_{\sss B}^{\sss\ S} \lb \rmE^{\sss A}_{\sss\ R} \G^{\sss R}_{\sss \ MS} - \dow_{\sss M}
  \rmE^{\sss A}_{\sss \ S} \rb \df x^{\sss M}, 
\end{equation} 
which has same information as $\G^{\sss R}_{\sss\ MS}$. So finally our system can be described by
the trio $\{\rmE^{\sss A}_{\sss\ M}, C^{\sss A}_{\sss \ MB}, A_\sM \}$ modded by
diffeomorphisms, gauge transformations, and $\rmS\rmO(d+1,1)$ rotations denoted by infinitesimal
parameters, 
\begin{equation}
  \p_\xi = \lbr \xi^{\sss M} \dow_{\sss M}, [\L_{\Sigma(\xi)}]^{\sss
    A}_{\sss\ B}, \L_{(\xi)} \rbr \in \rmT\cM_{(d+2)} \times \fs\fo(d+1,1)  \times \fg.  
\end{equation}
Here $\fs\fo(d+1,1)$ denotes the Lie algebra of $\rmS\rmO(d+1,1)$. $\p_\xi$ is given a Lie algebra
structure by defining a commutator on it,
\begin{equation}
  \p_{[\xi_1,\xi_2]} = \lB \p_{\xi_1}, \p_{\xi_2} \rB = \d_{\xi_1} \p_{\xi_2} = - \d_{\xi_2}
  \p_{\xi_1},
\end{equation} 
where,
\begin{align}
  \d_{\xi_1} \xi_2 &= \lie_{\xi_1} \xi_2 = - \lie_{\xi_2} \xi_1 = - \d_{\xi_2} \xi_1, \nn\\
  \d_{\xi_1} [\L_{\Sigma(\xi_2)}]^{\sss A}_{\sss \ B} &= \lie_{\xi_1} [\L_{\Sigma(\xi_2)}]^{\sss
                                                        A}_{\sss \ B} + [\L_{\Sigma(\xi_2)}]^{\sss A}_{\sss \ C} [\L_{\Sigma(\xi_1)}]^{\sss C}_{\sss \
                                                        B} - [\L_{\Sigma(\xi_1)}]^{\sss A}_{\sss \ C} [\L_{\Sigma(\xi_2)}]^{\sss C}_{\sss \ B}
                                                        - \lie_{\xi_2} [\L_{\Sigma(\xi_1)}]^{\sss A}_{\sss \ B}  \nn\\
                   &= - \d_{\xi_2} [\L_{\Sigma(\xi_1)}]^{\sss A}_{\sss \ B}, \nn\\
  \d_{\xi_1} \L_{(\xi_2)} &= \lie_{\xi_1} \L_{(\xi_2)} + \lB \L_{(\xi_2)}, \L_{(\xi_1)} \rB - \lie_{\xi_2} \L_{(\xi_1)} = - \d_{\xi_2} \L_{(\xi_1)}.
\end{align} 
Similarly the action of $\p_\xi$ (denoted by $\d_\xi$) on an arbitrary field $\vf$ (all indices
suppressed) obeys an algebra: $\lB \d_{\xi_1}, \d_{\xi_2} \rB \vf = \d_{[{\xi_1},{\xi_2}]} \vf$.
Under the action of $\p_{\xi}$ constituent fields vary as, 
\begin{align}\label{E:gauge_invariance_VR} 
\d_\xi
  \rmE^{\sss A}_{\sss \ M} &= \lie_\xi \rmE^{\sss A}_{\sss \ M} - [\L_{\Sigma(\xi)}]^{\sss A}_{\sss
    \ B} \rmE^{\sss B}_{\sss \ M}
  = \N_{\sss M} \xi^{\sss A} + \xi^{\sss N} \rmT^{\sss A}_{\sss \ NM} - [\nu_{\sss{\Sigma(\xi)}}]^{\sss A}_{\sss \ B} \rmE^{\sss B}_{\sss \ M}, \nn \\
  \d_\xi C^{\sss A}_{\sss \ MB} &= \lie_\xi C^{\sss A}_{\sss \ MB} + \N_{\sss M}
  [\L_{\Sigma(\xi)}]^{\sss A}_{\sss \ B}
  = \N_{\sss M} [\nu_{\sss{\Sigma(\xi)}}]^{\sss A}_{\sss \ B} + \xi^{\sss N} R_{\sss NM}{}^{\sss A}{}_{\sss B}, \nn \\
  \d_\xi A_{\sss M} &= \lie_\xi A_{\sss M} + \N_{\sss M} \L_{(\xi)} = \N_{\sss M} \nu_{\sss{(\xi)}}
  + \xi^{\sss N} F_{\sss NM}, 
\end{align} 
where $\xi^{\sss A} = \rmE^{\sss A}_{\sss \ M}\xi^{\sss M}$ and $\lie_\xi$ denotes Lie derivative
along $\xi^{\sss M}$. The covariant derivative $\N_{\sss M}$ is associated with all the connections
$\G^{\sss R}_{\sss \ MS}$, $C^{\sss A}_{\sss \ MB}$, $A_{\sss M}$, which acts on a general field
$\tensor{\vf}{^{\sss R}_{\sss S}^{\sss A}_{\sss B}}$ transforming in adjoint representation of the
gauge group as, \bee{\label{def_nabla} \N_{\sss M} \vf^{\sss R}{}_{\sss S}{}^{\sss A}{}_{\sss B} =
  \dow_{\sss M} \vf^{\sss R}{}_{\sss S}{}^{\sss A}{}_{\sss B} + \G^{\sss R}_{\sss\
    MN}\tensor{\vf}{^{\sss N}_{\sss S}^{\sss A}_{\sss B}} - \G^{\sss N}_{\sss \
    MS}\tensor{\vf}{^{\sss R}_{\sss N}^{\sss A}_{\sss B}} + C^{\sss A}_{\sss\ MC}
  \tensor{\vf}{^{\sss R}_{\sss S}^{\sss C}_{\sss B}} - C^{\sss C}_{\sss \ MB} \tensor{\vf}{^{\sss
      R}_{\sss S}^{\sss A}_{\sss C}} + \lB A_{\sss M} , \vf^{\sss R}{}_{\sss S}{}^{\sss A}{}_{\sss
    B} \rB, } and similarly on higher rank objects. In \cref{E:gauge_invariance_VR} we have
defined\footnote{\label{foot_scaledCP} By scaled we mean scaled with temperature: $\nu = \mu/\vq$, where $\mu$ is the
  chemical potential and $\vq$ is the temperature. Note that at this point these quantities are just
  introduced for computational convenience; and will get a physical meaning only in presence of a
  preferred symmetry data, e.g. when spacetime admits an isometry.}
\begin{align}\label{scaledCP}
  \text{Scaled gauge chemical potential:}&\qquad \nu_{\sss{(\xi)}} = \L_{(\xi)} + \xi^{\sss N} A_{\sss
    N}, \nn\\
  \text{Scaled spin chemical potential:}&\qquad [\nu_{\sss{\Sigma(\xi)}}]^{\sss A}_{\sss\ B} =
  [\L_{\Sigma(\xi)}]^{\sss A}_{\sss \ B} + \xi^{\sN} C^{\sss A}_{\sss\ NB},
\end{align}
associated with $\p_\xi$. We have also defined curvatures of all the constituent fields, \bea{
  \text{Gauge Field Strength:}&\qquad \bF = \df \bA + \bA \wedge \bA = \half F_{\sss MN} \df x^{\sss M} \wedge \df x^{\sss N}, \nn\\
  \text{Spacetime Curvature:}&\qquad \bm R^{\sss A}_{\sss \ B} = \df \bC^{\sss A}_{\sss\ B} + \bC^{\sss A}_{\sss\ C} \wedge \bC^{\sss C}_{\sss\ B} = \half R_{\sss MN}{}^{\sss A}_{\sss\ B}  \df x^{\sss M} \wedge \df x^{\sss N}, \nn\\
  \text{Spacetime Torsion:}&\qquad \mathbf T^{\sss A} = \df \mathbf E^{\sss A} + \bC^{\sss A}_{\sss\
    B} \wedge \mathbf E^{\sss B} = \half \rmT^{\sss A}_{\sss \ MN} \df x^{\sss M} \wedge \df x^{\sss
    N}.  } On can check that all these quantities $\nu_{\sss{(\xi)}}$,
$[\nu_{\sss{\Sigma(\xi)}}]^{\sss A}_{\sss \ B}$, $F_{\sss MN}$, $R_{\sss MN}{}^{\sss A}_{\sss \ B}$,
$\rmT^{\sss A}_{\sss\ MN}$ transform covariantly under the action of $\p_{\xi}$. It is interesting
to note that $C^\sA_{\ \sM\sB}$ transforms as a $\fs\fo(d+1,1)$ valued gauge field.  In terms of torsion
it is possible to give an exact expression for connections which we note for completeness, \bea{
  \G^{\sss R}_{\sss\ MS} &= \half \rmG^{\sss RN} \lb \dow_{\sss M} \rmG_{\sss NS} + \dow_{\sss S}
  \rmG_{\sss NM} - \dow_{\sss N} \rmG_{\sss MS}
  + \rmT_{\sss NMS} - \rmT_{\sss MSN} - \rmT_{\sss SMN} \rb, \nn\\
  C^{\sss A}_{\sss\ MB} &= \half \eta_{\sss BD} \rmE^{[{\sss D}|\sss S} \lB 2 \lb 2\dow_{[\sss S}
  \rmE^{{\sss A}]}{}_{{\sss M}]} - \rmT^{{\sss A}]}{}_{\sss SM} \rb + \rmE_{\sss CM} \rmE^{{\sss
      A}]\sss N} \lb 2\dow_{[\sss S} \rmE^{\sss C}_{\ {\sss N}]} - \rmT^{\sss C}_{\sss \ SN} \rb
  \rB.  } A physical theory on $\cM_{(d+2)}$ can be described by a partition function
$W[\rmE^{\sss A}_{\sss \ M}, C^{\sss A}_{\sss \ MB}, A_{\sss M}]$ which is a functional of Vielbein
and connections. Under an infinitesimal variation of the sources its response is captured by,
\bee{\label{consistent_variation_EinCar} \d W = \int \lbr \df x^{\sss M} \rbr \sqrt{|\rmG|} \Big(
  T^{\sss M}_{\sss \ \ A} \d \rmE^{\sss A}_{\sss \ M} + \Sigma^{\sss MA}_{\sss \ \ \ B} \d C^{\sss
    B}_{\sss \ MA} + J^{\sss M} \cdot \d A_{\sss M} \Big), } where $X\cdot Y = \Tr{XY}$ for
$X,Y\in \fg$ is the inner product on $\fg$. $T^{\sss MA}$ is the \emph{canonical energy-momentum
  tensor}, $\Sigma^{\sss MAB}$ is the \emph{spin current} (antisymmetric in its last two indices)
and $J^{\sss M}$ is the \emph{charge current}. Demanding the partition function to be invariant
under the action of $\p_\xi$ we can find the Ward
identities\footnote{\label{torsionless_EinCar_foot} Note that we can use the spin Ward identity to
  eliminate antisymmetric part of canonical EM tensor in the EM conservation equation. Doing this is
  particularly helpful in torsionless theories where the new EM conservation becomes,
  $\N_{\sss M} T^{\sss MN}_{(b)} = F^{\sss NM} \cdot J_{\sss M}$.  Here we have defined the
  symmetric \emph{Belinfante energy-momentum tensor},
  $T^{MN}_{(b)} = T^{(MN)} + 2\N_R \Sigma^{(MN)R} $. In this work however, we will mostly talk in
  terms of canonical EM tensor as this is the Noether charge corresponding to translations. Also it
  is well known that gravitational anomalies do not affect the canonical EM conservation
  \cite{Chang:1984ib}.  }  related to these currents, \bea{\label{EOM_noanom_EinCar}
  \underline{\N}_{\sss M} T^{\sss M}_{\sss \ \ N} &= \rmT^{\sss B}_{\sss \ NM} T^{\sss M}_{\sss \ \
    B} + R_{\sss NM}{}^{\sss A}_{\sss \ B} \Sigma^{\sss MB}_{\sss \ \ \ A}
  + F_{\sss NM}\cdot J^{\sss M}, \nn\\
  \underline{\N}_{\sss M} \Sigma^{\sss MAB} &=
  T^{[{\sss BA}]}, \nn\\
  \underline{\N}_{\sss M} J^{\sss M} &= 0, } where
$\underline{\N}_{\sss M} = \N_{\sss M} - \rmT^{\sss N}_{\sss \ NM}$.

\subsection{Null Backgrounds}

We are now ready to define null backgrounds. These kind of backgrounds and their Galilean
interpretation goes back to
\cite{Julia:1994bs,Christensen:2013rfa,Duval:1984cj,Bergshoeff:2014uea}. The idea of null
backgrounds is to somewhat tweak the procedure, so that we not only get the correct symmetries, but
also reproduce the required background field content after reduction. As we shall show, this even
allows us to add anomalies in odd dimensional null backgrounds which naively doesn't look possible.

We will call $\p_\xi$ a \emph{compatible symmetry data} if the scaled chemical potentials associated
with it defined in \cref{scaledCP} are identically zero. Now, a manifold $\cM_{(d+2)}$ along with
fields $\{\rmE^\sA_{\ \sM}, C^\sA_{\ \sM\sB}, A_{\sM}\}$ will be
called a \emph{null background} (or more formally a \emph{compatible Bargmann structure}) if it
admits a \emph{covariantly constant \underline{compatible} null isometry} generated by
$\p_{V} = \{ V^M \dow_M, [\L_{\Sigma(V)}]^\sA_{\ \sB}, \L_{(V)} \}$ i.e.,
\begin{enumerate}
	\item Action of $\p_V$ is an isometry, $\d_V \rmE^{\sss A}_{\sss \ M} = \d_V C^{\sss A}_{\sss \ MB}= \d_V A_{\sss M}  = 0$, 
	\item $V$ is null, $V^{\sss M} V_{\sss M} = 0$,
	\item $V$ is covariantly constant, $\N_{\sss M} V^{\sss N} = 0$, and
	\item $\p_V$ is compatible, $\nu_{\sss{(V)}} = V^{\sss M} A_{\sss M} + \L_{(V)} = 0$, $[\nu_{\sss{\Sigma(V)}}]^{\sss A}_{\sss \ B} = V^{\sss M} C^{\sss A}_{\sss \ MB} + [\L_{\Sigma(V)}]^{\sss A}_{\sss \ B} = 0$.
\end{enumerate}
Although this definition of null backgrounds is little different from \cite{Banerjee:2015hra}, one can check that it boils down to the same in torsionless limit. If we drop condition (4), i.e. compatibility, we would be left with the definition of \emph{Bargmann
  structures} \cite{Duval:1984cj} extended to Vielbein formalism. They have some nice properties,
\bee{\label{null_strengths_cond1}
  \rmT^{\sss A}_{\sss \ MN}V_{\sss A} = H_{\sss MN} \equiv 2\dow_{[\sM} V_{\sN]}, \qquad
  R_{\sss MN}{}^{\sss A}_{\sss  \ B} V_{\sA} = 0.
}
 Hence if we are interested in a torsionless theory, we would have to apply a dynamic constraint on
 $V$, which can be violated off-shell. Requirement of compatibility further imposes,
\bee{\label{null_strengths_cond}
	V^{\sss M}\rmT^{\sss A}_{\sss \ MN} = V^{\sss M} F_{\sss MN} = V^{\sss M} R_{\sss
          MN}{}^{\sss A}_{\sss  \ B} = 0, \qquad
        V^\sM\N_M \vf = \d_V \vf,
}
for any tensor $\vf$ transforming in an appropriate representation of $\fg$ and $\fs\fo(d+1,1)$ (all
indices suppressed). These restrictions are in some sense backbone of null backgrounds. First and foremost, they eliminate unphysical mass sources that would otherwise appear in the mass conservation law after reduction. Hints of it were originally  found by \cite{Banerjee:2014mka} in an attempt of naive null reduction of charged fluids. We would have more to say about it later. As we shall see, these restrictions also allow for anomalies on odd dimensional null backgrounds and forbid them in even dimensional ones. This is an important feature, if we are to reproduce physically realizable anomalies in Galilean theories in one lower dimension.

We demand that physical theories on null backgrounds (referred as null theories) are not invariant under action of any arbitrary $\p_\xi$ but only those which leave $\p_V$ invariant i.e. $[\p_{V},\p_{\xi}]=0$. This requirement ensures that there is no dynamics along the isometry even off-shell. The new partition function variation can be written following \cref{consistent_variation_EinCar} as,
\bee{\label{consistent_variation}
	\d W = \int \lbr \df x^{\sss M} \rbr \sqrt{|\rmG|} \lb T^{\sss M}_{\sss \ \ A} \d \rmE^{\sss A}_{\sss \ M} + \Sigma^{\sss MA}_{\sss \ \ \ B} \d C^{\sss B}_{\sss \ MA} + J^{\sss M}\cdot \d A_{\sss M} + \#_{\sss A} \d V^{\sss A} \rb.
}
Note the last term in this expression, which is valid since our restriction does not forbid us from varying $V^{\sss A}$. Astute reader might note that we could have absorbed that term into $T^{\sss MA}$ owing to the fact that $\d V^{\sss M} = 0$, but we have a better setup in mind. The conditions of null background along with the restrictions we have imposed, imply that null theories are invariant under the following set of current redefinitions,
\bee{\label{unphysic1}
	T^{\sss MA} \ra T^{\sss MA} + V^{\sss M} \q^{\sss A}_{1}, \qquad
	\Sigma^{\sss MAB} \ra \Sigma^{\sss MAB} + V^{\sss M} \q_2^{\sss AB}, 
	\qquad
	J^{\sss M} \ra J^{\sss M} + \q_3 V^{\sss M}, 
}
\bee{
	\#^{\sss A} \ra \#^{\sss A} - \q_1^{\sss A} + \q_4 V^{\sss A},
}
where $\q$'s are arbitrary scalars transforming in appropriate representations of $\fg$ and $\fs\fo(d+1,1)$. The Ward identities on null backgrounds will also be slightly modified compared to \cref{EOM_noanom_EinCar}\footnote{\label{torsionless_null_foot}
Following \cref{torsionless_EinCar_foot} one might wonder how the respective Belinfante EM conservation law looks like for null theories. Similar to non-null case one can use the spin conservation in EM conservation law, which will give $\N_{\sss M} \lb  T^{\sss MN}_{(b)} - \#^{[\sss N} V^{{\sss M}]}\rb =  F^{\sss NM}\cdot  J_{\sss M}$. One can show that the $\#^{\sss M}$ dependence can be removed by using $ T^{\sss MA}$ redefinition \cref{unphysic1}, after which one recovers the standard Belinfante conservation law (given in \cref{torsionless_EinCar_foot}) even for null theories. Belinfante EM tensor on the other hand is left with redefinition freedom, $ T^{\sss MN}_{(b)} \ra  T^{\sss MN}_{(b)} + \q_1 V^{\sss M} V^{\sss N}$. These were derived directly for a spinless null theory in \cite{Banerjee:2015uta}.
},
\bea{\label{EOM_noanom}
	\underline{\N}_{\sss M} T^{\sss M}_{\sss \ \ N}
	&=
	\rmT^{\sss B}_{\sss \ NM} T^{\sss M}_{\sss \ \ B} 
	+ R_{\sss NM}{}^{\sss A}_{\sss \ B} \Sigma^{\sss MB}_{\sss \ \ \ A}
	+ F_{\sss NM}\cdot J^{\sss M}, \nn\\
	\underline{\N}_{\sss M} \Sigma^{\sss MAB} &= 
	T^{[{\sss BA}]} + \#^{[\sss A} V^{{\sss B}]}, \nn\\
	\underline{\N}_{\sss M} J^{\sss M} &= 0,
}
One can check that these equations are invariant under redefinitions \cref{unphysic1}. To interpret the new $\#^{\sss A}$ term, note that spin (angular momentum) conservation are $\half (d+1)(d+2)$ equations. However as was pointed out in the introduction, after reduction system only respects $\half d(d-1)$ equations corresponding to rotations and $d$ equations corresponding to Galilean boosts. The job of $\#^{\sss A}$ is then to eliminate the remaining $(d+1)$ conservation equations, as it does. Practically it is best to fix an `off-shell gauge' $\d V^{\sss A} = 0$, which renders a new invariance in the spin current,
\bee{\label{unphysic2}
	\Sigma^{\sss MAB} \ra \Sigma^{\sss MAB} + \q_5^{{\sss M}[\sss A} V^{{\sss B}]},
}
and omits the remaining $(d+1)$ components of the spin conservation. Note that it will further
restrict $\p_\xi$ to obey $[\nu_{\sss{\Sigma(\xi)}}]^{\sss A}_{\sss \ B} V^{\sss B} = 0$. This point
onward we would assume every symmetry data $\p_\xi$ to satisfy these requirements, and would term
them \emph{$\p_V$ compatible symmetry data}. From this viewpoint, spin conservation in
\cref{EOM_noanom} must be true for some $\#^{\sss M}$, hence ruling out components involving
$\#^{\sss M}$ as they carry no information.

On null backgrounds using $\p_V$ we can also define some more `thermodynamic' variables associated
with $\p_\xi$ similar to \cref{scaledCP},
\begin{equation}\label{def_temperature}
	\text{Temperature:}\quad \vq_{\sss{(\xi)}} = - \frac{1}{\xi^{\sss N} V_{\sss N}}, \qquad
	\text{Scaled mass chemical potential:}\quad \vp_{\sss{(\xi)}} = - \frac{\xi^{\sss M} \xi_{\sM}}{2 \xi^{\sss N} V_{\sss N}},
\end{equation}
and using it we can define chemical potentials from scaled chemical potentials,
\bee{\label{chemicalpotentials}
	\mu_{\sss{(\xi)}} = \vq_{\sss{(\xi)}}\nu_{\sss{(\xi)}}, \qquad
	[\mu_{\sss{\Sigma(\xi)}}]^{\sss A}_{\sss \ B} = \vq_{\sss{(\xi)}}[\nu_{\sss{\Sigma(\xi)}}]^{\sss A}_{\sss \ B}, \qquad
	\mu_{\vp\sss{(\xi)}} = \vq_{\sss{(\xi)}}\vp_{\sss{(\xi)}}.
}
These abstract definitions will find use later.

\subsection{Null Reduction -- Newton-Cartan Backgrounds} \label{null_redn}

Having obtained the Ward identities in the null background language, it is now time to see what does
they imply for the Galilean theories. To do this we need to pick up a foliation
$\cM_{(d+2)} = S^1_{V} \times \cM^{\text{NC}}_{(d+1)}$ and compactify along the isometry direction
$V$. Following \cite{Banerjee:2015hra} we note that since $V$ is null, it is not possible to find a
unique such foliation without choosing a set of $\p_V$ compatible \emph{time data},
$\p_T = \{ T^{\sss M} \dow_{\sss M}, [\L_{\Sigma(T)}]^\sA_{\ \sB}, \L_{(T)} \}$. This is tantamount
to choosing a preferred Galilean frame of reference\footnote{\cite{Geracie:2015xfa} proposed a
  formalism for Galilean theories independent of the choice of frame. But on a closer look it would
  be clear that they have just discovered null backgrounds from a different perspective. The Ward
  identities as described by \cite{Geracie:2015xfa} are just the null background Ward identities
  with slight rearrangement; we give a comparison in \cref{Geracie:2015xfa}.}. Having chosen $\p_T$
we can define such a foliation as $\cM_{(d+2)} = S^1_{V} \times \bbR_T \times \cM^T_{(d)}$, where we
identify $\cM^{\mathrm{NC}}_{(d+1)} = \bbR_T \times \cM^T_{(d)}$ as degenerate \emph{Newton-Cartan}
(NC) manifold. We define \emph{null reduction} as this choice of foliation and subsequent
compactification.

\paragraph*{Newton-Cartan Structure:} Using $\p_T$ we can define a null field orthonormal to $V$ as, \bee{\label{defn_barV}
  \bar V^{\sss M}_{(T)} = \vq_{\sss{(T)}} T^{\sss M} + \mu_{\vp\sss{(T)}} V^{\sss M},} such that
$\bar V^{\sss M}_{(T)} \bar V_{(T)\sss M} = 0$, and $\bar V^{\sss M}_{(T)} V_{\sss M} = -1$. Here
$\vq_{\sss{(T)}}$, $\mu_{\vp\sss{(T)}}$ have been defined in
\cref{def_temperature,chemicalpotentials}. Without loss of generality we choose a basis
on $\cM_{(d+2)}$, $x^{\sss M} = \{ x^{\sss\sim}, x^\mu \}$ such that
$\p_V = \{\dow_{\sss \sim}, 0 , 0 \}$. On the other hand on $\bbR^{(d+1,1)}$ we choose a basis
$x^\sA = \{ x^-, x^+ , x^a \}$ such that $V = \dow_-$ and $\bar V_{(T)} = \dow_+$. At this stage we
choose a specific representation of $\eta_{\sss AB}$, $\rmE^{\sss A}_{\sss \ M}$ and
$\rmE_{\sss A}^{\sss \ M}$ compatible with the mentioned basis, \bee{ \eta_{\sss AB}
  = \begin{pmatrix} 0 & -1 & 0 \\ -1 & 0 & 0 \\ 0 & 0 & \d_{ab} \end{pmatrix}, \qquad \rmE^{\sss
    A}_{\sss \ M} = \begin{pmatrix}
    1	& - B_\mu \\
    0	& n_\mu \\
    0 & e^a_{\ \mu}
	\end{pmatrix}, \qquad
	\rmE_{\sss A}^{\sss \ M} = \begin{pmatrix}
		1	& 0 \\
		B_\nu v^\nu 	& v^\mu \\
		B_\nu e^{\ \nu}_{a}	& e^{\ \mu}_{a}
	\end{pmatrix},
}
such that,
\bee{
	n_\mu v^\mu = 1, \qquad 
	e^{\ \mu}_a n_\mu = 0, \qquad
	e_{\ \mu}^a v^\mu = 0, \qquad
	e^{a}_{\ \mu} e^{\ \mu}_b = \d^a_{\ b}, \qquad
	v^\mu n_\nu + e^{\ \mu}_a e^{a}_{\ \nu}  = \d^\mu_{\ \nu}.
}
This can be identified as \emph{Newton-Cartan} (NC) structure. We can also define the NC degenerate metric by,
\bee{
	h_{\mu\nu} = e^a_{\ \mu} e^b_{\ \nu} \d_{ab}, \qquad
	h^{\mu\nu} = e_a^{\ \mu} e_b^{\ \nu} \d^{ab}.
}
Since there is no non-degenerate metric on $\cM^{\mathrm{NC}}_{(d+1)}$, raising/lowering of $\mu,\nu\ldots$ indices is not permitted. However $a,b\ldots$ indices can be raised/lowered using $\d_{ab}$. NC Vielbein $e^a_{\ \mu}$ is not a `square matrix' and hence does not furnish an invertible map between tensors on $\cM^{\mathrm{NC}}_{(d+1)}$ and $\bbR^{(d)}$. It however can be used to project tensors on $\cM^{\mathrm{NC}}_{(d+1)}$ to tensors on $\bbR^{(d)}$, and tensors on $\bbR^{(d)}$ to `spatial tensors' on $\cM^{\mathrm{NC}}_{(d+1)}$,
\bee{
	e^a_{\ \mu} X^\mu = X^a, \qquad e_{a}^{\ \mu} Y_\mu = Y_a, \qquad
	X^a e_{a}^{\ \mu} = h^{\mu}_{\ \nu} X^\nu, \qquad Y_a e^{a}_{\ \mu} = h^\nu_{\ \mu} Y_\nu,
}
where $h^{\mu}_{\ \nu} = h^{\mu\r}h_{\s\nu}$. The compatibility of null isometry switches off many components of the connections $\G^{\sss M}_{\sss \ \sim N }$, $\G^{\sss M}_{\ \mu \sss\sim}$, $C^{\sss A}_{\sss \ \sim B}$, $C^{\sss A}_{\ \mu -}$, $C^+_{\ \mu \sss B}$ and $A_\sim$. Remaining non-zero components can be determined to be,
\bea{
	C^{-}_{\ \mu a} &= c_{\mu a}, \qquad
	C^{a}_{\ \mu +} = c_{\mu}^{\ a}, \qquad
	\G^{\sss \sim}_{\ \mu\nu} = c_{\mu\nu} - \Ndot_{\mu} B_{\nu}, \nn\\
	\G^{\l}_{\ \mu\nu}
	&= 
	v^\l \dow_{\mu} n_{\nu}
	+ \half h^{\l\s} \lb \dow_\mu h_{\s\nu} + \dow_\nu h_{\s\mu} - \dow_\s h_{\mu\nu} \rb
	+ n_{(\mu} \O_{\nu)\s} h^{\l\s}
	+ \half \lb 
		e_{a}^{\ \l} \rmT_{\ \mu\nu}^{a}
		- 2 e_{a(\nu} \rmT^a_{\ \mu)\s} h^{\l\s}
	\rb, \nn\\
	C^a_{\ \mu b} 
	&= 
	\half n_{\mu} \O_{b}{}^a
	+ \half \eta_{bd} e^{[d|\nu}  \lB
		2 \lb 2\dow_{[\nu} e^{a]}_{\ \mu]} - \rmT^{a]}_{\ \nu\mu}  \rb
		+ e_{c\mu} e^{a]\s} \lb 2\dow_{[\nu} e^c_{\ \s]} - \rmT^c_{\ \nu\s}  \rb
	\rB.
}
Here we have defined spacetime dependence of frame velocity $c_{\mu\nu} = h_{\s\nu} \Ndot_{\mu}v^\s$
in terms of which frame vorticity is given by $\O_{\mu\nu} = 2 c_{[\mu\nu]}$. We call a time data (reference frame) $\p_T$ to be \emph{globally inertial} if $c_{\mu\nu} = 0$. We choose the connections on $\cM^{\mathrm{NC}}_{(d+1)}$ to be $\G^\l_{\ \mu\nu}$, $C^a_{\ \mu b}$ and $A_\mu$, and denote associated covariant derivative by $\Ndot_\mu$,  acting on a general field $\tensor{\vf}{^\r_\s^a_b}$ transforming in adjoint representation of the gauge group as,
\bee{
	\Ndot_\mu \tensor{\vf}{^\r_\s^a_b} = \dow_\mu \tensor{\vf}{^\r_\s^a_b}
	+ \G^\r_{\ \mu\nu}\tensor{\vf}{^\nu_\s^a_b}
	- \G^\nu_{\ \mu\s}\tensor{\vf}{^\r_\nu^a_b}
	+ C^a_{\ \mu c} \tensor{\vf}{^\r_\s^c_b}
	- C^c_{\ \mu b} \tensor{\vf}{^\r_\s^a_c}
	+ \lB A_\mu , \tensor{\vf}{^\r_\s^a_b} \rB,
}
and similarly on higher rank objects. Action of $\Ndot_\mu$ on NC structure can be found to be,
\bee{
	\Ndot_\mu n_\nu = 0, \quad
	\Ndot_\mu e_a^{\ \nu} = 0, \quad
	\Ndot_\mu h^{\r\s} = 0, \quad
	\Ndot_\mu h_{\nu\r} = - 2 c_{\mu(\nu} n_{\r)}, \quad
	\Ndot_\mu e^a_{\ \nu} = - n_\nu c_{\mu}^{\ a}.
}
One can check that $\Ndot_\mu, \G^\l_{\ \mu\nu}$ agrees with the most generic NC covariant
derivative and connection written down in \cite{Geracie:2015xfa}. One can also perform the reduction
of curvatures. Surviving components of gauge field strength are $F_{\mu\nu}$ which act as NC gauge
field strength. Similarly surviving components of torsion are spatial torsion $\rmT^{a}_{\ \mu\nu}$,
`mass torsion' $\rmT_{+\mu\nu} = - \rmT^{-}_{\ \mu\nu}$ and temporal torsion $H_{\mu\nu} = -
\rmT^{+}_{\ \mu\nu}$. Finally we have the surviving components of curvature,
\bee{
	\tensor{R}{_{\mu\nu}^a_+} = 2\dow_{[\mu} \tensor{c}{_{\nu]}^a} + 2C^a_{\ [\mu|b}\tensor{c}{_{\nu]}^b}, \qquad
	R_{\mu\nu}{}^{a}{}_b = 2\dow_{[\mu} C^a_{\ \nu]b} + 2C^a_{\ [\mu|c}C^c_{\ \nu] b},
}
which act as NC temporal and spatial curvatures respectively. Both curvatures can also be combined into a full NC curvature, 
\bee{
	\tensor{R}{_{\mu\nu}^{\r}_\s}
	= 
	\tensor{e}{_a^\r}  \lb
		\tensor{R}{_{\mu\nu}^{a}_+} n_\s
		+ \tensor{R}{_{\mu\nu}^{a}_b} \tensor{e}{^b_\s}
	\rb.
}
We define raised NC volume element,
\bee{
	\ve_\uparrow^{\mu\nu\ldots} =  \bar V_M \e^{M\mu\nu\ldots} = - \e^{-\mu\nu\ldots}.
}
Again since volume element is defined with all indices up, and there is no lowering operation, the corresponding Hodge dual $\ast_\uparrow$ gives a map from differential forms to completely antisymmetric contravariant tensor fields. It is also possible to define a lowered volume element, but we would not require it for our purposes. More details on NC volume forms and Hodge duals can be found in \cref{forms}.

\paragraph*{Conserved Currents and Ward Identities:} Now we need to decompose the currents in this basis,
\bee{\label{redn_constitutive}
	T^{\sss M}_{\sss \ \ A} = \begin{pmatrix}
		\times		& \times		& \times \\
		- \r^\mu	& - \e^\mu	& p^\mu_{\ a}
	\end{pmatrix}, \quad
	J^{\sss M} = \begin{pmatrix}
		\times \\
		j^\mu
	\end{pmatrix}, \quad
	\Sigma^{\sss \sim AB} = \times, \quad
	\Sigma^{\mu \sss  AB} = \begin{pmatrix}
		0 & \times & \times \\
		\times	& 0 & \t^{\mu b} \\
		\times	& - \t^{\mu a}  & \s^{\mu ab}
	\end{pmatrix}.
}
Here we have denoted unphysical components by $\times$ which can be eliminated by redefinitions \cref{unphysic1,unphysic2}. We identify $\r^\mu$ as mass current, $\e^\mu$ as energy current, $p^{\mu a}$ as momentum current, $\t^{\mu a}$ as temporal spin current, $\s^{\mu ab}$ as spatial spin current, and finally $j^\mu$ as charge current. We can also project the $\mu$ index in these currents onto $\bbR^{(d)}$ to get corresponding `spatial currents'. On the other hand we define various densities as the projection of these currents along $ n_\mu$,
\bee{
	\r = n_\mu \r^\mu, \quad
	\e = n_\mu \e^\mu, \quad
	p^a = n_\mu p^{\mu a}, \quad
	\t^a = n_\mu \t^{\mu a}, \quad
	\s^{ab} = n_\mu \s^{\mu ab}, \quad
	q = n_\mu j^\mu.
} 
In terms of these, physical components of Ward identities \cref{EOM_noanom} can be expressed as,
\bea{\label{EOM_gal}
	\text{Mass Cons. (Continuity):}&\qquad \underline{\Ndot}_\mu \r^\mu = 0, \nn\\
	\text{Energy Cons. (Time Translation):}&\qquad \underline{\Ndot}_\mu \e^{\mu}
	=
	[\text{power}]
	- p^{\mu a} c_{\mu a}, \nn\\
	\text{Momentum Cons. (Translations):}&\qquad \underline{\Ndot}_\mu p^{\mu}_{\ a} 
	=
	[\text{force}]_a
	- \r^{\mu} c_{\mu a}, \nn\\
	\text{Temporal Spin Cons. (Galilean Boosts):}&\qquad \underline{\Ndot}_\mu \tensor{\t}{^{\mu}_{a}}
	= 
	\half \lb \r_{a} - p_a \rb, \nn\\
	\text{Spatial Spin Cons. (Rotations):}&\qquad  \underline{\Ndot}_\mu \s^{\mu ab} = 
	p^{[ba]}
	+ 2 \t^{\mu[a} c_\mu^{\ b]}, \nn\\
	\text{Charge Cons. (Gauge Transformations):}&\qquad  \underline{\Ndot}_\mu j^\mu = 0,
}
where $\underline{\Ndot}_\mu = \Ndot_\mu + v^\nu H_{\nu\mu} - e_a^{\ \nu}\rmT^a_{\ \nu\mu}$. These are the (non-anomalous) conservation laws of a Galilean theory with spin current. We have mentioned above what are the conserved quantities (and what is the underlying symmetry). 
The temporal conservation equation, which is slightly less familiar, is akin to the Milne boost Ward identity of torsionless case, which states that spatial mass current must equal the momentum density (look e.g. \cite{Jensen:2014aia} and follow references therein). Here $[\text{power}]$ and $[\text{force}]_a$ are power and force densities due to background fields,
\bea{\label{power.work}
	[\text{power}] &= - v^\nu \lb
		H_{\nu\mu} \e^{\mu}
		+ \rmT_{+\nu\mu} \r^{\mu}
		+ \rmT^a_{\ \nu\mu} p^{\mu}_{\ a}
		+ \tensor{R}{_{\mu\nu}^a_+} \tensor{\t}{^{\mu}_a}
		+ R_{\nu\mu ab} \s^{\mu ba}
		+ F_{\nu\mu}\cdot j^\mu
	\rb, \nn\\
	[\text{force}]_a &= e_a^{\ \nu} \lb
		H_{\nu\mu} \e^{\mu}
		+ \rmT_{+\nu\mu} \r^{\mu}
		+ \rmT^a_{\ \nu\mu} p^{\mu}_{\ a}
		+ \tensor{R}{_{\mu\nu}^a_+} \tensor{\t}{^{\mu}_a}
		+ R_{\nu\mu ab} \s^{\mu ba}
		+ F_{\nu\mu}\cdot j^\mu
	\rb,
}
which act as energy and momentum sources respectively. The terms coupling to $c_{\mu a}$ in \cref{EOM_gal} are due to
the chosen Galilean frame (time data) not being globally inertial and hence causes pseudo-power, pseudo-force and pseudo-torque.

One could have taken a slightly different approach to get these Ward identities and performed null reduction at the level of partition function \cref{consistent_variation} itself,
\bee{\label{NR_PF}
	\d W = \int \lbr \df x^\mu \rbr \sqrt{|\g|} \lb 
		\r^{\mu} \d B_{\mu}
		- \e^{\mu} \d n_{\mu}
		+ p^{\mu}_{\ a} \d e^a_{\ \mu}
		+ 2\t^{\mu a} \d c_{\mu a} 
		+ \s^{\mu a}{}_{b} \d C^b_{\ \mu a} 
		+ j^\mu\cdot \d A_\mu
	\rb,
}
where $\g_{\mu\nu} = h_{\mu\nu} + n_\mu n_\nu$ and $\g = \det \g_{\mu\nu} = - \rmG$. Symmetry data
$\p_\xi$ breaks up in NC basis as, 
\begin{equation}
  \p^{\text{NC}}_\xi = \lbr \quad \L_{\rmM(\xi)} \equiv -\xi^{\sss \sim}, \quad \xi^\mu, \quad [\L_{\t(\xi)}]_a \equiv [\L_{\Sigma(\xi)}]^{-}{}_{a},
  \quad [\L_{\s(\xi)}]^a_{\ b} \equiv [\L_{\Sigma(\xi)}]^{a}_{\ b}, \quad \L_{(\xi)} \quad \rbr,
\end{equation}
Variation of various constituent fields under the action of $\p_{\xi}^{\text{NC}}$ (also denoted as $\d_\xi$) can be obtained via null reduction\footnote{
	Note that fixing $V^{\sss M}$ or $V^{\sss A}$ is not a `gauge fixing', as transformations
        shifting these are not part of our symmetries on null backgrounds. On the other hand fixing $\bar V_{(T)}^{\sss A}$ is a gauge fixing which can be violated off-shell. If we fix this gauge even off-shell we would miss the corresponding temporal spin conservation equation.
},
\begin{align}\label{NC_BF_var}
	\d_\xi B_\mu &= 
	\lie_\xi B_{\mu}
	+ \dow_\mu \L_{\rmM(\xi)}
	+ [\L_{\t(\xi)}]_{a} e^a_{\ \mu} 
	= \dow_\mu \nu_{\sss\rmM(\xi)}
	+ \xi^\nu \rmT_{+\nu\mu}
	- \xi^\nu c_{\mu\nu}
	+ [\nu_{\t\sss{(\xi)}}]_{a} e^a_{\ \mu} \nn\\
	\d_\xi n_\mu &= 
	\lie_\xi n_\mu = \dow_\mu \xi^+
	- \xi^\nu H_{\nu\mu} \nn \\
	\d_\xi e^a_{\ \mu} &= 
	\lie_\xi e^a_{\ \mu}
	- [\L_{\s(\xi)}]^a_{\ b} e^b_{\ \mu}
	- [\L_{\t(\xi)}]^{a} n_{\mu} 
	=
	\Ndot_\mu \xi^a
	+ \xi^+ c_\mu{}^a
	+ \xi^\nu \rmT^a_{\ \nu\mu}
	- [\nu_{\s\sss{(\xi)}}]^a_{\ b} e^b_{\ \mu}
	- [\nu_{\t\sss{(\xi)}}]^{a} n_{\mu} \nn\\
	\d_\xi c_{\mu a} 
	&= 
	\lie_\xi c_{\mu a}
	+ \lb\dow_\mu [\L_{\t(\xi)}]_{a} - C^b_{\ \mu a} [\L_{\t(\xi)}]_{b} \rb
	+ [\L_{\s(\xi)}]^b_{\ a} c_{\mu b} 
	= \Ndot_{\mu} [\nu_{\t {\sss(\xi)}}]_{a} 
	+ [\nu_{\s\sss{(\xi)}}]^b_{\ a} c_{\mu b} 
	- \xi^\nu R_{\nu\mu +a} \nn\\
	\d_\xi C^a_{\ \mu b}
	&=
	\lie_\xi C^a_{\ \mu b}  
	+ \lb 
		\dow_\mu [\L_{\s(\xi)}]^a_{\ b} 
		+ C^a_{\ \mu c} [\L_{\s(\xi)}]^c_{\ b} 
		- C^c_{\ \mu b} [\L_{\s(\xi)}]^a_{\ c} 
	\rb 
	= \Ndot_{\mu} [\nu_{\s\sss{(\xi)}}]^a_{\ b} + \xi^\nu R_{\nu\mu \ b}^{\ \ a}, \nn\\
	\d_\xi A_\mu &= 
	\lie_\xi A_{\mu}
	+ \dow_\mu \L_{(\xi)} + [A_\mu, \L_{(\xi)}]
	= 
	\Ndot_\mu \nu_{\sss{(\xi)}} + \xi^\nu F_{\nu\mu}.
\end{align}
Looking at these expressions we can identify $\L_{\sss\rmM(\xi)}$ as continuity
parameter, $\xi^\mu$ the spacetime translation parameter, $[\L_{\t(\xi)}]_{a}$ as Galilean boost
parameter, $[\L_{\s(\xi)}]^{a}_{\ b}$ as rotation parameter, and $\L_{(\xi)}$ as gauge parameter. It is further noteworthy that $\xi^+ = n_\mu \xi^\mu$ and $\xi^a = e^a_{\ \mu} \xi^\mu$ serve as time translation and space translation parameters respectively. Demanding invariance of \cref{NR_PF} under all these parameters one can recover Ward identities \cref{EOM_gal}. One can compare these results to those of \cite{Geracie:2015xfa}.

In the first equation of (\ref{NC_BF_var}) we have defined \emph{scaled \underline{total} mass
  chemical potential} associated with $\p_\xi$ as
$\nu_{\sss\rmM(\xi)} = \L_{\rmM(\xi)} + \xi^\mu B_\mu = \xi^\sM \bar V_{(T)\sM}$. It is differs from
the scaled mass chemical potential $\vp_{(\xi)}$ defined in \cref{def_temperature}, by a `kinetic'
part, $\nu_{\sss\rmM(\xi)} = \vp_{\sss(\xi)} - \frac{1}{2\vq_{\sss(\xi)}} \bar V_{(\xi)}^a \bar V_{(\xi)a} $. Following
\cref{chemicalpotentials} we can also define total mass chemical potential as,
$\mu_{\sss\rmM(\xi)} = \vq_{\sss(\xi)}\nu_{\sss\rmM(\xi)} = \mu_{\vp\sss(\xi)} - \frac{1}{2} \bar V_{(\xi)}^a \bar V_{(\xi)a}$.\footnote{Although we will not be using it in this work, it is interesting to differentiate
  in two types of mass chemical potentials. Consider that our system has a preferrend symmetry data $\p_U$. Naively $\mu_{\vp}$ corresponds to the first law of thermodynamics written
  in terms of internal energy $E$, while $\mu_{\sss\rmM}$ corresponds to the first law in terms of
  total energy $E_{\text{tot}} = E + \half R u^a u_a$ (where $u^a = \bar V_{(U)}^a$; subscripts $\scriptstyle(U)$ have
  been dropped),
  \begin{equation}
    \df E = \vq \df S + \mu_\vp \df R + [\mu_{\sss\Sigma}]^\sB_{\ \sA} \df [Q_\Sigma]^\sA_{\ \sB} +
    \mu \cdot \df Q, \quad
    \df E_{\text{tot}} = \vq \df S + \mu_{\sss\rmM} \df R + u^a \df (R u_a) +
    [\mu_{\sss\Sigma}]^\sB_{\ \sA} \df [Q_\Sigma]^\sA_{\ \sB} +
    \mu \cdot \df Q.
  \end{equation}
  When working with total energy as a thermodynamic variable, the thermodynamics becomes frame
  dependent and first law has a term corresponding to work done due to momentum density $R u_a$ as
  well. Notation used here can be found in \cite{Banerjee:2015hra}.
}

Finally we would like to note that mass being exactly conserved is a consequence of compatibility. Otherwise the respective conservation equation would look something like,
\begin{align}\label{mass.sources}
	- \underline{\Ndot}_\mu \r^\mu
	&=
	\rmT^{\sss A}_{\sss \ \sim M} T^{\sss M}_{\sss \ \ A} 
	+ R_{\sss \sim M}{}^{\sss A}_{\ \sss B} \Sigma^{\sss MB}_{\sss\ \ \ A}
	+ F_{\sss \sim M}\cdot J^{\sss M}\nn\\
	&=
	T^{\sss M}_{\sss \ \ A} \lb 
		\rmE^{\sss B}_{\sss \ M} [\nu_{\sss\Sigma(V)}]^{\sss A}_{\sss \ B} 
		- \rmE^{\sss A}_{\sss \ N}\N_{\sss M} V^{\sss N}
	\rb
	- \Sigma^{\sss M B}_{\sss \ \ \ A} \N_{\sss M} [\nu_{\sss\Sigma(V)}]^{\sss A}_{\sss \ B}
	- J^{\sss M}\cdot \Ndot_{\sss M} \nu_{\sss{(V)}}.
\end{align}
One can clearly see that $\N_{\sss M} V^{\sss N}$, $[\nu_{\sss\Sigma(V)}]^{\sss A}_{\sss \ B}$ and $\nu_{\sss(V)}$ source this conservation. One of the prime reasons for imposing compatibility is to get rid of these mass sources.

Comparing our analysis to the torsionless case of \cite{Banerjee:2015hra} one would note that authors there also imposed a `$T$-redefinition' invariance in the theory, which leads to Galilean boost transformation upon reduction. Note that on defining $\bar\p^\mu = [\L_{\t(\xi)}]^{a} e_{a}^{\ \mu}$, our Galilean boost transformation,
\bee{
	\d_\xi \big\vert_{\bar\p} B_\mu = 
	\bar\p_{\mu}, \qquad
	\d_\xi \big\vert_{\bar\p} v^\mu = \bar\p^\mu, \qquad
	\d_\xi \big\vert_{\bar\p} h_{\mu\nu} = - 2 \bar\p_{(\mu} n_{\nu)},
}
boils down to (infinitesimal) $T$-redefinition transformation of \cite{Banerjee:2015hra}. Hence for us imposing $T$-redefinition is redundant. Actually even for \cite{Banerjee:2015hra}, imposing $T$-redefinition was redundant, as the authors noted that the corresponding Ward identity is trivially satisfied for theories obtained by null reduction. It was helpful however to have this transformation there, because Galilean currents are not boost invariant and there was no non-trivial inherent symmetry of the partition function to keep track of these transformations.

\section{Galilean Gauge and Spin Anomalies} \label{anomaly_inflow}

In the previous section we used null reduction to obtain Ward identities for a Galilean theory with non-trivial spin current. Now we would like to take this a step ahead and ask -- how these identities modify in presence of a gauge and gravitational anomalies. We would give away the suspense right away, because the following story is quite technical.  As one would expect, the gauge anomaly in null theory translates to gauge anomaly in Galilean theory as well, while the gravitational anomaly manifests itself purely through spatial spin conservation. Other $4$ out of $6$ conservation laws in \cref{EOM_gal} remain non-anomalous. In formulation of anomalies in Cartan language it is not surprising; it is known that gravitational anomaly acts as Lorentz anomaly in this formalism and only violates spin conservation \cite{Chang:1984ib}. What is surprising is that we did not find any anomalies in temporal spin conservation (or correspondingly Milne boost Ward identity). We do not claim that this anomaly cannot be introduced by other means or that we are not missing anything, but the fact that number of anomaly coefficients in our treatment and that of a relativistic theory match exactly (in fact they both are determined by the same anomaly polynomial), gives us some confidence in our results. 

\subsection{Anomaly Inflow on Einstein-Cartan Backgrounds}

In relativistic theories anomaly inflow has been by far the best way to understand gauge and gravitational anomalies \cite{Callan:1984sa}. We would like to take a step back and first describe the anomaly inflow mechanism for generic Einstein-Cartan theories. The extension to null theories will then be more transparent. A good discussion on anomaly inflow for torsionless relativistic theories can be found in \S 2 of \cite{Jensen:2012kj}. We consider that our manifold of interest $\cM_{(d+2)}$ lives on the boundary of a \emph{bulk} manifold $ \cB_{(d+3)}$. Bulk coordinates are denoted with a bar, and we choose a basis $x^{\bar{\sss M}} = \{ x^{\perp}, x^{\sss M} \}$, where $x^{\perp}$ corresponds to depth into the bulk. All the field content $\rmE^{\bar {\sss A}}_{\ \bar {\sss M}}$, $A_{\bar {\sss M}}$, $C^{\bar {\sss A}}_{\ \bar {\sss M} \bsss{B}}$ is extended down into the bulk with the requirement that all $\perp$ components vanish at the boundary.

Now we keep our theory of interest on $\cM_{(d+2)}$, whose generating functional $W_{\cM}$ is not necessarily invariant under symmetries of the theory -- i.e. is anomalous. In the bulk we keep some theory with generating functional $W_{\cB}$, which is invariant under all symmetries up to some non-trivial boundary terms. The full theory described by $W = W_{\cM} + W_{\cB}$ is assumed to be invariant under all symmetries. It is actually this non-trivial boundary term in $W_{\cB}$ which induces anomaly in the boundary theory, hence the name \emph{anomaly inflow}. Note that in absence of anomalies $W_{\cB} = 0 \Ra W = W_\cM$ which was discussed in last section. Let us assume for now that we have figured out such a $W_{\cB}$, and parametrize its infinitesimal variation as\footnote{Note that $\rmS\rmO(d+1,1)$ transformations leave the flat metric $\eta_{\sss AB}$ invariant, hence it can commute freely through variations.},
\bem{\label{W_bulk}
	\d W_{\cB}
	= 
	\int \lbr \df x^{\bsss M} \rbr \sqrt{|\rmG_{(d+3)}|} \lb 
		\rmT^{\bsss M\bsss A}_{\rmH} \d \rmE_{\bsss A \bsss M}
		+ \Sigma^{\bsss M\bsss A\bsss B}_{\rmH} \d C_{\bsss B\bsss M \bsss A}
		+ \rmJ^{\bsss M}_{\rmH}\cdot \d A_{\bsss M}
	\rb \\
	+ \int \lbr \df x^{\sss M} \rbr \sqrt{|\rmG|} \lb \rmT_{\rmB\rmZ}^{\sss MA} \d \rmE_{\sss AM} 
		+ \Sigma^{\sss MAB}_{\rmB\rmZ}\d C_{\sss BMA} + \rmJ_{\rmB\rmZ}^{\sss M} \cdot \d A_{\sss M} \rb.
}
It is generally known that $W_{\cB}$ is topological and hence does not depend on the metric/Vielbein, but we keep it here just for the sense of generality; we will see the respective terms vanishing when we put in the allowed expression for $W_{\cB}$. The \emph{Hall currents} in the bulk must be manifestly symmetry covariant by definition of $W_{\cB}$. The boundary \emph{Bardeen-Zumino currents} on the other hand are symmetry non-covariant. Variation of $W_\cM$ will generate the \emph{consistent currents} which due to anomaly are not symmetry covariant either,
\bee{
	\d W_\cM
	= 
	\int \lbr \df x^{\sss M} \rbr \sqrt{|\rmG|} \lb T_{\text{cons}}^{\sss MA} \d \rmE_{\sss AM} + \Sigma^{\sss MAB}_{\text{cons}}\d C_{\sss BMA} + J_{\text{cons}}^{\sss M}\cdot \d A_{\sss M} \rb.
}
Since the full partition function $W$ should be symmetry invariant, we can read out the symmetry covariant, \emph{covariant currents} in the boundary,
\bee{
	T^{\sss MA} = T_{\text{cons}}^{\sss MA} + \rmT_{\rmB\rmZ}^{\sss MA}, \qquad
	\Sigma^{\sss MAB} = \Sigma_{\text{cons}}^{\sss MAB} + \Sigma_{\rmB\rmZ}^{\sss MAB}, \qquad
	J^{\sss M} = J^{\sss M}_{\text{cons}} + \rmJ^{\sss M}_{\rmB\rmZ}.
}
Demanding $W$ to be invariant under all symmetries of the theory, we will get the anomalous Ward identities for these currents,
\bea{\label{EOM_usual}
	\underline{\N}_{\sss M} T^{\sss M}_{\sss \ \ N}
	&=
	\rmT^{\sss A}_{\sss \ NM} T^{\sss M}_{\sss \ \ A} 
	+ R_{\sss NM}{}^{\sss A}_{\sss \ B} \Sigma^{\sss MB}_{\sss \ \ \ A}
	+ F_{\sss NM}\cdot J^{\sss M}
	+ \rmT^{\perp}_{\rmH}{}_{\sss N}, \nn\\
	\underline{\N}_{\sss M} \Sigma^{\sss MAB} &= T^{[{\sss BA}]} + \Sigma_{\rmH}^{\perp\sss  AB}, \nn\\
	\underline{\N}_{\sss M} J^{\sss M} &= \rmJ_{\rmH }^\perp.
}
We verify that the bulk Hall currents source anomaly in the boundary theory. On the other hand Hall currents themselves must satisfy the non-anomalous Ward identities \cref{EOM_noanom_EinCar} in the bulk, which would be trivial if $W_{\cB}$ is chosen properly. Now depending on the field content of the theory one would have to construct the most generic allowed $W_{\cB}$ and read out from there the Hall currents. This would determine the most generic anomalies that can occur in the respective theory which can be modeled using anomaly inflow. In notation of differential forms $W_{\cB}$ is given by integration of a full rank form $\bI^{(d+3)}$,
\bee{\label{Wbulk}
	W_{\cB} = \int_{ \cB_{(d+3)}} \bI^{(d+3)}.
}
Requirement that its variation should be symmetry invariant up to a boundary term can be recasted into the requirement that $\bcP^{(d+4)} = \df \bI^{(d+3)}$ should be symmetry invariant. $\bcP^{(d+4)}$ is called the anomaly polynomial, which encodes all the non-trivial information about anomaly. It is evident that $\bcP^{(d+4)}$ needs to be closed, symmetry invariant, and should not be expressible as exterior derivative of a symmetry invariant form. For example on usual backgrounds (not null), $\bcP^{(2n+4)}$ is given by the Chern-Simons anomaly polynomial $\bcP^{(2n+4)}_{\rmC\rmS }$ for even dimensional boundary theories, and no such term is possible in odd dimensions. $\bcP^{(2n+4)}_{\rmC\rmS }$ is a `polynomial' made out of Chern classes of $\bF$ and Pontryagin classes of $\bR$. Look e.g. \cite{Jensen:2012kj} for more details.

\subsection{Anomaly Inflow on Null/Newton-Cartan Backgrounds} \label{null_inflow}

Now we come back to our case of interest -- null backgrounds. Above procedure goes more or less through, except that bulk $ \cB_{(d+3)}$ is now required to possess a compatible null isometry $\p_V$, which translates itself to a compatible null isometry on the boundary $\cM_{(d+2)}$ since all the $\perp$ components vanish. Variation of $W_{\cB}$ in \cref{W_bulk} remains unchanged under a $\p_V$ compatible variation, except all the currents now follow redefinitions specified in \cref{unphysic1,unphysic2}. Consequently we can find the anomalous Ward identities for null backgrounds,
\bea{\label{EOM}
	\underline{\N}_{\sss M} T^{\sss M}_{\sss \ \ N}
	&=
	\rmT^{\sss A}_{\sss \ NM} T^{\sss M}_{\sss \ \ A} 
	+ R_{\sss NM}{}^{\sss A}_{\sss \ B} \Sigma^{\sss MB}_{\sss \ \ \ A}
	+ F_{\sss NM}\cdot J^{\sss M}
	+ \rmT^\perp_{\rmH \sss N}, \nn\\
	\underline{\N}_{\sss M} \Sigma^{\sss MAB} &= T^{[{\sss BA}]} + \Sigma_{\rmH }^{\perp \sss AB} + \#^{[\sss A} V^{{\sss B}]}, \nn\\
	\underline{\N}_{\sss M} J^{\sss M} &= \rmJ_{\rmH}^\perp,
}
for some $\#^{\sss M}$. These are same as non-null identities except that just like non-anomalous
case some components of spin current conservation have been discarded using spin current
redefinition \cref{unphysic2}. Physical components of these laws can be expressed after reduction as
anomalous Galilean conservation laws,
\bea{\label{EOM_anom_NC}
	\text{Mass Cons. (Continuity):}&\qquad \underline{\Ndot}_\mu \r^\mu = \uprho^\perp_{\rmH }, \nn\\
	\text{Energy Cons. (Time Translation):}&\qquad \underline{\Ndot}_\mu \e^{\mu}
	=
	[\text{power}]
	- p^{\mu a} c_{\mu a}
	+ \upepsilon^\perp_{\rmH }, \nn\\
	\text{Momentum Cons. (Translations):}&\qquad \underline{\Ndot}_\mu p^{\mu}_{\ a} 
	=
	[\text{force}]_a
	- \r^{\mu} c_{\mu a} + \mathrm p^\perp_{\rmH a}, \nn\\
	\text{Temporal Spin Cons. (Galilean Boosts):}&\qquad \underline{\Ndot}_\mu \t^{\mu a} 
	= 
	\half \lb \r^a - p^a\rb
	+ \uptau_{\rmH }^{\perp a}, \nn\\
	\text{Spatial Spin Cons. (Rotations):}&\qquad  \underline{\Ndot}_\mu \s^{\mu ab} = 
	p^{[ba]}
	+ 2 \t^{\mu[a} c_{\mu}^{\ b]}
	+ \upsigma^{\perp ab}_{\rmH }, \nn\\
	\text{Charge Cons. (Gauge Transformations):}&\qquad  \underline{\Ndot}_\mu j^\mu = \mathrm j_{\rmH }^\perp,
}
where we have decomposed the Hall currents as,
\begin{equation}
	\rmT^{\perp}_{\rmH \sss A} = \begin{pmatrix}
		- \uprho_{\rmH }^\perp	& - \upepsilon_{\rmH }^\perp	& \mathrm p^\perp_{\rmH a}
	\end{pmatrix}, \qquad
	\rmJ^{\perp}_{\rmH } = \mathrm j_{\rmH }^\perp, \qquad
	\Sigma^{\perp AB}_{\rmH } = \begin{pmatrix}
		0 & \times & \times \\
		\times	& 0 & \uptau^{\perp b}_\rmH \\
		\times	& - \uptau^{\perp a}_\rmH  & \upsigma^{\perp ab}_{\rmH }
              \end{pmatrix}.  
\end{equation} 
We hence see that in principle anomaly inflow can destroy all the conservation laws. It is now the
form of $\bcP^{(d+4)}$ which will determine how many of these anomalies are permissible and in what
dimensions.

On even dimensional $(d=2n)$ null backgrounds the allowed anomaly polynomial takes the usual
Chern-Simons structure of relativistic theories $\bcP^{(2n+4)} = \bcP_{\rmC\rmS }^{(2n+4)}$ which is
made up of Chern classes of $\bF$ and Pontryagin classes of $\bR$. Note however that neither of
$\bF,\bR$ have a leg along $V$, hence $\bcP_{\rmC\rmS }^{(2n+4)}$ is identically zero. The
corresponding $\bI_{\rmC\rmS }$ might still have a leg along $V$ since
$\i_V \bA ,\i_V \bC^{\sss A}_{\sss \ B} \neq 0$ for a general null theory. But one can check that
the corresponding $\perp$ components of (dual) Hall currents again have no leg along $V$ and hence
the Ward identities become non-anomalous. This suggests that we cannot get anomalies in an even
dimensional null theory, and hence odd dimensional Galilean theories are anomaly free.

At this point we would like to point out some subtle differences from the analysis of
\cite{Jensen:2014hqa}. In the cited reference author does not impose compatibility of the isometry,
and hence $\bF$, $\bR$ does have a leg along $V$. This results in anomalous conservation laws that
crucially depend on $\nu_{\sss{(V)}}$, $[\nu_{\sss{\Sigma(V)}}]^{\sss A}_{\sss \ B}$ -- additional
fields which are otherwise switched off by compatibility. As we mentioned in the introduction, we
have chosen to switch off these fields as they serve as a `mass source' in the Galilean theory, and
we do not see these sources in non-relativistic theories that occur in nature.

Now we shift our attention to the more interesting case of odd dimensional ($d=2n-1$) null
backgrounds. One can check that with the field content at hand, it is not possible to naively
construct an anomaly polynomial. Following \cite{Banerjee:2015hra} however, we note that we can
remedy this problem by introducing auxiliary \emph{time data} $\p_T$ that was used to perform null
reduction in \cref{null_redn}. Using the corresponding $\bar V_{(T)}$ defined in \cref{defn_barV},
we can write the only allowed anomaly polynomial, \bee{\label{E:anomalypol_odd} \bcP^{(2n+3)} =
  \bm{{\bar V}}_{(T)} \wedge \bcP^{(2n+2)}_{\rmC\rmS }, } where
$\bm{{\bar V}}_{(T)} = \bar V_{(T){\sss M}} \df x^{\sss M}$. Although this expression has an
explicit dependence on $\p_T$, one can show that it is invariant under any arbitrary redefinition of
$\p_T$. This follows from the fact that change in $\bar V_{(T)\sss M}$ does not have any leg along
$V$, due to normalization property
$\d (\bar V_{(T)\sss M} V^{\sss M}) = V^{\sss M} \d\bar V_{(T)\sss M} = 0$. For this reason we drop
the subscript ${\scriptstyle (T)}$ from $\bar V_{(T)}$ this point onward. Readers should convince
themselves that there are no more terms which can be written in the anomaly polynomial. However, we
have a problem; anomaly polynomial \cref{E:anomalypol_odd} is not exact, \bee{ \bcP^{(2n+3)} = - \df
  \lb \bm{{\bar V}} \wedge \bI^{(2n+1)}_{\rmC\rmS } \rb + \df \bm{{\bar{V}}} \wedge
  \bI^{(2n+1)}_{\rmC\rmS }.  } Hence for $\bI^{(2n+2)}$ to be well defined, the second term must
vanish. In general however it does not as $\bI^{(2n+1)}_{\rmC\rmS }$ does have a leg along $V$. It
hence forces us to choose \emph{transverse gauge} for the null isometry generated $\p_V$, i.e., \bee{ \L_{(V)}
  = [\L_{\Sigma(V)}]^{\sss A}_{\sss \ B} = 0, } which ensures $\bI^{(2n+1)}_{\rmC\rmS }$ does not
have any leg along $V$. Some comments are due; different choices of $\p_V$ represent different null
theories, as we are not allowed to perform transformations which alter these (we demanded the
partition function to be invariant under $\p_V$ preserving transformations). Hence the statement is
that not all such null theories can be anomalous; only null theories with transverse null isometry
can exhibit anomalies.  Note that in conventional null reduction, one generally chooses
$\p_V = \{ \dow_{\sss \sim} , 0, 0 \}$ which by definition satisfies the transversality
requirement. Modulo this subtlety we can find, \bee{ \bI^{(2n+2)} = - \bm{{\bar V}} \wedge
  \bI^{(2n+1)}_{\rmC\rmS }.  } Computing its variation one can find \emph{Hall} and
\emph{Bardeen-Zumino} currents defined in \cref{W_bulk}, \bee{\nn \rmT^{\bsss M \bsss A}_{\rmH } =
  0, \qquad \star_{(2n+2)}\bm\Sigma_{\rmH }^{\bsss A\bsss B} = \bm{{\bar V}} \wedge \frac{\dow
    \bcP_{\rmC\rmS }^{(2n+2)}}{\dow \bR_{\bsss B\bsss A}}, \qquad \star_{(2n+2)}\bm\rmJ_{\rmH } =
  \bm{{\bar V}} \wedge \frac{\dow \bcP_{\rmC\rmS }^{(2n+2)}}{\dow \bF}, }
\bee{\label{hallcurrent_def} \rmT^{\sss MA}_{\rmB\rmZ } = 0, \qquad \star\bm\Sigma_{\rmB\rmZ }^{\sss
    AB} = \bm{{\bar V}} \wedge \frac{\dow \bI_{\rmC\rmS }^{(2n+1)}}{\dow \bR_{\sss BA}}, \qquad
  \star\bm\rmJ_{\rmB\rmZ } = - \bm{{\bar V}} \wedge \frac{\dow \bI_{\rmC\rmS }^{(2n+1)}}{\dow \bF}.
} We verify that $T^{\bsss M \bsss A}_{\rmH }, T^{\sss MA}_{\rmB\rmZ }$ vanish. It immediately
follows that mass, energy and momentum conservation are non-anomalous. Also the (Milne) boost Ward
identity stays non-anomalous as matrix indices of $\Sigma_{\rmH }^{\sss MAB}$ comes from
$\bR^{\sss A}_{\sss \ B}$ which have a zero contraction along $V$. Again this follows from
compatibility of isometry, and is not true for considerations of \cite{Jensen:2014hqa}, which is why
they find a Milne anomaly. These statements can be recasted as, 
\begin{equation}\label{NC_Hall1} 
  \uprho^\perp_{\rmH} = \upepsilon^\perp_{\rmH } = \mathrm p^{\perp a}_{\rmH } = \uptau^{\perp a}_{\rmH } = 0, 
\end{equation} 
which follows directly
from null reduction. The only laws that get anomalous are hence spin and charge
conservation. Explicit expressions for their Hall currents follow from reduction,
\begin{equation}\label{NC_Hall2} 
  \mathrm j_{\rmH }^\perp = - \ast_\uparrow \lB \frac{\dow \bm\fp^{(2n+2)}}{\dow \bF}\rB, \qquad 
  \upsigma_{\rmH }^{\perp ab} = - \ast_\uparrow \lB \frac{\dow \bm\fp^{(2n+2)}}{\dow \bR_{ba}} \rB.  
\end{equation} 
Here we have formally denoted $\bcP_{\rmC\rmS }^{(2n+2)}$ as $\bm\fp^{(2n+2)}$ after
reduction; the distinction is purely notational. $\ast_\uparrow$ is the Hodge dual associated with
raised Newton-Cartan volume element $\bm\ve_\uparrow$; look \cref{forms} for more details. Putting
back \cref{NC_Hall1,NC_Hall2} into \cref{EOM_gal} we can get the anomalous Ward identities for
Galilean theories.

Before closing this section, we would like to make some due comments on even dimensional case. One
might worry that we can use $\p_T$ to define anomalies in even dimensions as well. However one can
check that only possible symmetry-covariant anomaly polynomial we can write involving $\p_T$ is,
\bee{\label{E:anomalypol_even2} \bcP^{(2n+4)} = \bV \wedge \bm{{\bar V}} \wedge
  \bcP^{(2n+2)}_{\rmC\rmS }, } where $\bV = V_{\sss M} \df x^{\sss M}$. This anomaly polynomial is
however not an exact form, \bee{ \bcP^{(2n+4)} = \df \lb \bV \wedge \bm{{\bar V}} \wedge
  \bI^{(2n+1)}_{\rmC\rmS } \rb - \bH \wedge \bm{{\bar V}} \wedge \bI^{(2n+1)}_{\rmC\rmS } + \bV
  \wedge \df \bm{{\bar V}} \wedge \bI^{(2n+1)}_{\rmC\rmS }.  } The last term can be removed just
like before by going to transverse gauge, but the second last term cannot. We hence see that the
current formalism does not allow for anomalous even dimensional null theories. From this point
onward we will assume our null background to be odd dimensional, and hence set $d = 2n-1$.

With this we conclude our discussion on generic anomalous Galilean theories. Using the construction
of null backgrounds, we have found a set of conservation laws which determine the dynamics of these
theories in terms of a set of currents. These laws have already been well explored in literature,
but the fact that they follow by trivially choosing a basis in a higher dimensional null theory is
to be appreciated. Going along lines of \cite{Geracie:2015xfa}, it appears to us that null
backgrounds are the true `covariant' and `frame independent' formalism of Galilean physics, which
appear pretty natural from a 5 dimensional perspective. We refer the reader to
\cref{Geracie:2015xfa} for more comments on these issues.

All of the results presented here are in Newton-Cartan notation, which is the natural covariant
prescription for Galilean physics. In \cref{non_cov} we present some of our results in conventional
non-covariant notation, for the benefit of readers who are not comfortable with the Newton-Cartan
language. Even otherwise, seeing the results in non-covariant form might help us relate it better to
day to day physics, where we are used to viewing time and space separately.

\section{Anomalous Galilean Hydrodynamics} \label{hydro}

In previous sections we have obtained the anomalous conservation laws for a null/Galilean theory
with non-zero spin current. Here we want to study these theories in hydrodynamic limit -- near
equilibrium effective description of any quantum system. Before going to that let us make some
general comments about hydrodynamics on Einstein-Cartan backgrounds.  We start by picking up a
collection of \emph{hydrodynamic fields} which can be exactly solved for using the equations of
motion of the theory. Since there is an equation of motion for each symmetry data, we choose
hydrodynamic fields to be a set of symmetry data\footnote{We drop the subscript ${\scriptstyle(U)}$
  for $\p_{U}$ and hope that it will be clear by context.}
$\p_U = \{ U^{\sss M}, [\L_{\Sigma}]^{\sss A}_{\sss \ B}, \L \}$. The \emph{fluid} (hydrodynamic
system) is characterized by conserved currents $T^{\sss MA}$, $\Sigma^{\sss MAB}$, $J^{\sss M}$
written as the most generic tensors made out of hydrodynamic fields $\p_U$ and background sources
$\rmE^{\sss A}_{\sss \ M}$, $C^{\sss A}_{\sss \ MB}$, $A_{\sss M}$, arranged in a derivative
expansion. These are known as \emph{constitutive relations} of the fluid. Near equilibrium
assumption of hydrodynamics implies that derivatives of quantities are small compared to quantities
themselves, which allows for proper truncation of the derivative expansion as dictated by the
cause. Dynamics of these constitutive relations in turn is governed by the conservation laws
\cref{EOM_usual}.  These constitutive relations are further subjected to the second law of thermodynamics,
i.e. the requirement of an entropy current $S^{\sss M}$ such that
$\underline{\N}_{\sss M} S^{\sss M} \geq 0$, whenever equations of motion are satisfied. This
requirement imposes various constraints on the constitutive relations, and the job of hydrodynamics
is to monitor these constraints. Having done so, one can in principle plug these constitutive
relations back into the equations of motion and solve for exact `configurations' of hydrodynamic
fields, which is not in the scope of hydrodynamics. A nice and modern review of relativistic
hydrodynamics can be found in \S 1 of \cite{Haehl:2015pja}.

Another notion which is inherent to any statistical system is \emph{equilibrium}. Equilibrium is the steady state of hydrodynamics, when the fluid has come in terms with the background and has aligned itself accordingly. In this state, the fluid can be described by a partition function $W^{eqb}$ written purely in terms of background data, and equations of motion are trivially satisfied. Equilibrium is generally defined by a collection of symmetry data $\p_K = \{ K^{\sss M}, [\L_{\Sigma(K)}]^{\sss A}_{\ \sss B}, \L_{(K)} \}$ which acts as an isometry on the background. For our constitutive relations to be physical, we will need to ensure that on introducing $\p_K$ they trivially satisfy the equations of motion \cref{EOM_usual}. 

Please note that $\p_U$ is a set of variables we have picked up to solve the system; like in any field theory we could do an arbitrary field redefinition of $\p_U$ without changing the physics. This is known as \emph{hydrodynamic redefinition freedom}. By convention $\p_U$ is defined to agree with $\p_K$ in equilibrium at zero derivative order (this goes into definition of fluid velocity, temperature and chemical potential in equilibrium), which fixes a huge amount of this freedom. Further fixing of this freedom can be dealt in various different ways, which takes the name of hydrodynamic frames (find more thorough discussion on these frames for null fluids in \cite{Banerjee:2015hra}). Here we would work in the so called \emph{equilibrium frame} where $\p_U = \p_K$ exactly in equilibrium, not just at zero derivative order. Note that this does not fix the freedom completely, we can still perturb this relation with anything that vanishes in equilibrium. For now we conclude that on setting $\p_U = \p_K$ i.e. on promoting $\p_U$ to an isometry, the constitutive relations should identically satisfy the equations of motion.

It was noted in \cite{Loganayagam:2011mu} for relativistic fluids that it is helpful to remove the clause `whenever equations of motion are satisfied' from the second law requirement and upgrade it to an off-shell statement \cite{raey}, which for us will read,
\bem{\label{offshell_2ndLaw}
	\underline{\N}_{\sss M} S^{\sss M} 
	+ U^{\sss N} \lb 
		\underline{\N}_{\sss M} T^{\sss M}_{\sss \ \ N}
		- \rmT^{\sss A}_{\sss \ NM} T^{\sss M}_{\sss \ \ A} 
		- R_{\sss NM}{}^{\sss A}_{\sss \ B} \Sigma^{\sss MB}_{\sss \ \ \ A}
		- F_{\sss NM}\cdot J^{\sss M}
	\rb \\
	+ [\nu_{\sss\Sigma}]_{\sss BA} \lb \underline{\N}_{\sss M} \Sigma^{\sss MAB} - T^{[{\sss BA}]} - \Sigma_{\rmH }^{\perp \sss AB} \rb
	+ \nu\cdot \lb \underline{\N}_{\sss M} J^{\sss M} - \rmJ_{\rmH }^\perp \rb \geq 0.
}
This statement is slightly different from what was considered for torsionless case in
\cite{Loganayagam:2011mu}, but we verify its equivalence with theirs in \cref{belinEC}. 

Now we come back to null fluids -- fluids on null backgrounds. On null backgrounds, hydrodynamic
data $\p_U$ needs to be compatible with $\p_V$, i.e. $[\p_V,\p_{U}] = 0$ and
$[\nu_{\sss\Sigma}]^{\sss A}_{\sss \ B} V^{\sss B} = 0$. This makes sense because (1) resulting
constitutive relations must follow the null isometry, (2) not all components of the spin
conservation in \cref{EOM} are physical. Further, constitutive relations are allowed to depend on
$\p_V$ as well. One can check that upon making these tweaks, off-shell second law
\cref{offshell_2ndLaw} remains unchanged. We can now go back and study the most generic constitutive
relations for null fluids, which has been thoroughly considered in \cite{Banerjee:2015hra} for a
charged spinless torsionless null fluid with $\rmU(1)$ anomalies up to leading order in
derivatives. In this work however, we are only interested in the sector of hydrodynamics that is
governed and is completely determined by the anomalies\footnote{In relativistic hydrodynamics it is
  known \cite{Jensen:2012kj} that there are certain coefficients which appear as independent
  constants in naive derivative expansion, but can be fixed in terms of anomaly coefficients
  appearing at higher derivative orders by demanding consistency of euclidean vacuum. Similar
  constants have also showed up for Galilean fluids in \cite{Jensen:2014ama,Banerjee:2015hra}, but
  their connection to anomaly is not yet clear. Here however we do not consider these
  contributions.}. To accomplish this task in relativistic fluids, \cite{Jensen:2013kka} (see also
\cite{Jensen:2013rga}) proposed a mechanism based on transgression forms, which allows us to
`integrate' the anomalous equations of motion \cref{EOM_usual} and directly figure out the anomalous
contribution to constitutive relations. We will attempt to extend this construction to null fluids.

\subsection{Anomalous Null Fluids}

We start by defining hydrodynamic shadow gauge field and spin connection,
\bee{
	\bm{{\hat A}} = \bA + \mu \bV, \qquad
	\bm{{\hat C}}{}^{\sss A}_{\sss \ B} = \bC^{\sss A}_{\sss \ B} + [\mu_{\sss\Sigma}]^{\sss A}_{\sss \ B} \bV,
}
where $\mu$, $[\mu_{\sss\Sigma}]^\sA_{\ \sB}$ are gauge and spin chemical potentials associated with
$\p_U$ defined in \cref{chemicalpotentials}. 
One can check that both $\p_U,\p_V$ are compatible with this new gauge field and spin connection, i.e.,
\bee{\nn
	\hat\nu = U^{\sss M} \hat A_{\sss M} + \L = 0, \qquad
	[\hat\nu_{\sss\Sigma}]^{\sss A}_{\sss \ B} = U^{\sss M} \hat C^{\sss A}_{\sss \ MB} + [\L_{\Sigma}]^{\sss A}_{\sss \ B} = 0,
}
\bee{
	\hat\nu_{\sss (V)} = V^{\sss M} \hat A_{\sss M} = 0, \qquad
	[\hat\nu_{\sss\Sigma(V)}]^{\sss A}_{\sss \ B} = V^{\sss M} \hat C^{\sss A}_{\sss \ MB} = 0.
}
Recall that we have chosen $ \L_{(V)} = [\L_{\Sigma(V)}]^{\sss A}_{\sss \ B} = 0$ to be able to
define anomalies. We define the operation $(\bm{{\hat{ \ }}})$ as $\bm{{\hat\mu}} = \bm\mu \lb\bA
\ra \bm{{\hat A}}, \bC^\sA_{\ \sB} \ra\bm{{\hat C}}{}^\sA_{\ \sB}\rb$. One can check that the hatted field strengths also follow the null background conditions \cref{null_strengths_cond,null_strengths_cond1}. We would like to import one result from transgression machinery without proof (see \S 11 of \cite{Nakahara:2003nw} for more details), which implies that,
\bee{\label{trans_expr}
	\bI^{(2n+1)}_{\rmC\rmS } - \bm{{\hat I}}^{(2n+1)}_{\rmC\rmS } = \bcV_{\bcP_{\rmC\rmS }}^{(2n+1)} + \df\bcV_{\bI_{\rmC\rmS }}^{(2n)},
}
where,
\bee{
	\bcV_{\bcP_{\rmC\rmS }}^{(2n+1)} = 
	\frac{\bV}{\bH} \wedge \lb \bcP^{(2n+2)}_{\rmC\rmS } - \bm{{\hat\cP}}^{(2n+2)}_{\rmC\rmS } \rb, \qquad
	\bcV_{\bI_{\rmC\rmS }}^{(2n)} =
	\frac{\bV}{\bH} \wedge \lb \bI^{(2n+1)}_{\rmC\rmS } - \bm{{\hat I}}^{(2n+1)}_{\rmC\rmS } \rb.
}
One can check that these quantities are well defined. We argue that fluid in equilibrium configuration can be described by a (bulk + boundary) partition function $W^{eqb} = \cW_{\cB}^{eqb} + W^{eqb}_\cM$ which has been discussed in preceding sections. Away from equilibrium however, the system is described by an effective action $S = S_{\cB} + S_\cM$ which boils down to $\cW^{eqb}$ is equilibrium. We claim that appropriate $S_{\cB}$ to generate anomalous sector of null hydrodynamics is\footnote{It was argued by \cite{Haehl:2013hoa} that while this effective action is appropriate to give solutions to the off-shell second law of thermodynamics, minimization of this action with respect to dynamic fields does not give the correct dynamics. To get the correct dynamics we need to further modify this action in Schwinger-Keldysh formalism, which we do not touch upon here.},
\bee{
	S_{\cB} = W_{\cB} + \int_{ \cB_{(2n+2)}} \bm{{\bar V}} \wedge \bm{{\hat I}}^{(2n+1)}_{\rmC\rmS }
	= - \int_{ \cB_{(2n+2)}} \bm{{\bar V}} \wedge \lb \bI^{(2n+1)}_{\rmC\rmS } - \bm{{\hat I}}^{(2n+1)}_{\rmC\rmS } \rb.
}
In equilibrium ($\p_U = \p_K$) and on choosing transverse gauge for $\p_K$ ($\L_{(K)} =
[\L_{\Sigma(K)}]^\sA_{\ \sB} = 0$) the added piece vanishes, as it does not have any leg along $V$,
and we will recover the equilibrium partition function. Using \cref{trans_expr} we can decompose $S_{\cB}$ as,
\bee{\label{S_bulk_decom}
	S_{\cB} = 
	\int_{ \cB_{(2n+2)}} \bcV_{\bcP}^{(2n+2)} 
	+ \int_{\cM_{(2n+2)}} \bcV_{\bI}^{(2n+1)},
}
where we have identified,
\bea{\label{transgression}
	\bcV_{\bcP}^{(2n+2)} &= \frac{\bV}{\bH} \wedge \lb \bcP^{(2n+3)} - \bm{{\hat\cP}}^{(2n+3)} \rb
	= - \bm{{\bar V}} \wedge \frac{\bV}{\bH} \wedge \lb \bcP_{\rmC\rmS }^{(2n+2)} - \bm{{\hat\cP}}_{\rmC\rmS }^{(2n+2)} \rb, \nn\\
	\bcV_{\bI}^{(2n+1)} &= \frac{\bV}{\bH} \wedge \lb \bI^{(2n+2)} - \bm{{\hat I}}^{(2n+2)} \rb
	= \bm{{\bar V}} \wedge \frac{\bV}{\bH} \wedge \lb \bI_{\rmC\rmS }^{(2n+1)} - \bm{{\hat I}}_{\rmC\rmS }^{(2n+1)} \rb.
}
The bulk term in \cref{S_bulk_decom} is manifestly symmetry invariant, and full $S$ is symmetry
invariant by definition, hence if we decompose $S_{\cM} = S_{\text{n-a}} + S_{\cM,\text{anom}}$ with first piece being totally symmetry invariant we can infer,
\bee{\label{S_anom_bdy}
	S_{\cM,\text{anom}} = - \int_{ \cB_{(2n+2)}} \bm{{\bar V}} \wedge \bm{{\hat I}}^{(2n+1)}_{\rmC\rmS }.
}
$S_{\cM,\text{anom}}$ will generate anomalous sector of consistent currents. On the other hand for full effective action we will be left with $S = S_{\text{anom}} + S_{\text{n-a}}$ where,
\bee{
	S_{\text{anom}} = \int_{ \cB_{(2n+2)}} \bcV_{\bcP}^{(2n+2)}.
}
$S_{\text{anom}}$ will generate anomalous sector of covariant currents.

\paragraph*{Constitutive Relations:}

In light of our discussion above, we should be able to generate anomalous sector of covariant
currents by varying $S_{\text{anom}}$. We will get,
\bem{
	\d S_{\text{anom}}
	= 
	\int_{ \cB_{(2n+2)}}\bigg(
		\d\bA \wedge \cdot \star_{(2n+2)}\bm\rmJ_{\rmH }
		- \d \bm{{\hat A}} \wedge \cdot \star_{(2n+2)}\bm{{\hat \rmJ}}_{\rmH } \\
		+ \d\bC^{\bsss A}_{\ \bsss B} \wedge \star_{(2n+2)}\bm\Sigma_{\rmH }{}^{\bsss B}_{\ \bsss A}
		- \d \bm{{\hat C}}^{\bsss A}_{\ \bsss B} \wedge \star_{(2n+2)}\bm{{\hat\Sigma}}_{\rmH }{}^{\bsss B}_{\ \bsss A} 
	\bigg)\\
	+ \int_{\cM_{(2n+1)}} \lb 
		\d\bA \wedge \cdot \star\bJ_{\bcP}
		+ \d\bC^{\sss A}_{\sss \ B} \wedge \star\bm\Sigma_{\bcP}{}^{\sss B}_{\sss \ A}
		+ \d\bV \wedge \star\bE_{\bcP}
	\rb,
}
where we have defined,
\bea{\label{anom_currents}
	\star\bE_{\bcP}
	&= \frac{\dow \bcV^{(2n+2)}_{\bcP}}{\dow \bH}
	= \frac{\bV}{\bH^{\wedge 2}} \wedge \Bigg[ 
		\bm{{\hat \cP}}^{(2n+3)} - \bcP^{(2n+3)}
		- \bH \wedge \star_{(2n+2)} \lb 
			\mu\cdot \bm{{\hat \rmJ}}_{\rmH }
			+  [\mu_{\sss\Sigma}]^{\sss A}_{\sss \ B} \bm{{\hat\Sigma}}_{\rmH }{}^{\sss B}_{\sss \ A}
		\rb
	\Bigg], \nn\\
	\star\mathbf{\Sigma}_{\bcP}{}^{\sss A}_{\sss \ B}
	&= \frac{\dow \bcV^{(2n+2)}_{\bcP}}{\dow \bR^{\sss B}_{\sss \ A}}
	= \frac{\bV}{\bH} \wedge \star_{(2n+2)} \lb \bm\Sigma_{\rmH }{}^{\sss A}_{\sss \ B} - \bm{{\hat\Sigma}}_{\rmH }{}^{\sss A}_{\sss \ B}\rb, \nn\\
	\star\bJ_{\bcP}
	&= \frac{\dow \bcV^{(2n+2)}_{\bcP}}{\dow \bF}
	= \frac{\bV}{\bH} \wedge \star_{(2n+2)}\lb \bm\rmJ_{\rmH } - \bm{{\hat \rmJ}}_{\rmH } \rb,
}
and Hall currents have been defined in \cref{hallcurrent_def}. Since $S_{\text{anom}}$ is invariant under symmetries by construction, we can find a set of Bianchi identities these currents must follow,
\bea{\label{anom_EOM}
	\underline{\N}_{\sss M} (T^{\sss M}_{\ \ \sss N})_\rmA
	&= 
	\rmT^{\sss A}_{\sss \ NM} (T^{\sss M}_{\sss \ \ A})_{\rmA}
	+ R_{\sss NM}{}^{\sss A}_{\sss \ B} (\Sigma^{\sss MB}_{\sss \ \ \ A})_\rmA
	+ F_{\sss NM} (J^{\sss M})_{\rmA}
	- V_{\sss N} \lb
		\mu \cdot \hat \rmJ_{\rmH}^\perp
		+ [\mu_{\sss\Sigma}]_{\sss AB} \hat \Sigma_{\rmH }^{\perp \sss  BA}
	\rb, \nn\\
	\underline{\N}_{\sss M} (\Sigma^{\sss MAB})_{\rmA} &= 
	(T^{[{\sss BA}]})_{\rmA}
	+ \Sigma_{\rmH }^{\perp \sss  AB} - \hat \Sigma_{\rmH }^{\perp \sss  AB} + \#^{[\sss A} V^{{\sss B}]}, \nn\\
	\underline{\N}_{\sss M} (J^{\sss M})_{\rmA} &= \rmJ_{\rmH }^\perp - \hat \rmJ_{\rmH }^\perp,
}
where we have defined anomalous `class' of constitutive relations,
\bee{\label{anomT}
	(T^{\sss MA})_{\rmA} = E_{\bcP}^{\sss M} V^{\sss A}, \qquad
	(\Sigma^{\sss MAB})_{\rmA} = \Sigma^{\sss MAB}_{\bcP}, \qquad
	(J^{\sss M})_{\rmA} = J^{\sss M}_{\bcP}.
}
One can check that on plugging in $\p_U=\p_K$, hatted Hall currents vanish as they do not have any
leg along $K$ and $V$ simultaneously. Consequently the Bianchi identities \cref{anom_EOM} reduce to equations of motion \cref{EOM}; in other words the currents $(T^{\sss MA})_{\rmA}$, $(\Sigma^{\sss MAB})_{\rmA}$, $(J^{\sss M})_{\rmA}$ identically satisfy the equations of motion in equilibrium configuration, as required.

We would like to remind the reader that $\bar V$ was added as an arbitrary choice of frame and anomaly polynomial was invariant under $\p_T$ redefinition which shifts $\bar V$. The currents we have constructed should then also be invariant under $\p_T$ redefinition. One can check that under a $\p_T$-redefinition currents in \cref{anom_currents} shift by a closed form. By definition currents always have this ambiguity, hence we do not change the physics. In hydrodynamics most natural choice of $\p_T$ to define anomalies is $\p_U$.

\paragraph*{Adiabaticity and Entropy Current:} To claim that the currents we have constructed are physical, we must find a $(S^{\sss M})_{\rmA}$ which satisfies the off-shell second law \cref{offshell_2ndLaw}. Anomalous sector is bound to be parity violating, implying that no scalar expression can be ensured positive definite. This turns \cref{offshell_2ndLaw} into a more stringent condition,
\bem{\label{adiabaticity}
	\underline{\N}_{\sss M} (S^{\sss M} )_\rmA
	+ U^{\sss N} \Big[
		\underline{\N}_{\sss M}( T^{\sss M}_{\sss \ \ N})_\rmA
		- \rmT^{\sss A}_{\sss \ NM} (T^{\sss M}_{\sss \ \ A} )_\rmA
		- R_{\sss NM}{}^{\sss A}_{\sss \ B} (\Sigma^{\sss MB}_{\sss \ \ \ A})_\rmA
		- F_{\sss NM}\cdot (J^{\sss M})_\rmA
	\Big] \\
	+ [\nu_{\sss\Sigma}]_{\sss BA} \Big[ \underline{\N}_{\sss M} (\Sigma^{\sss MAB})_\rmA - (T^{[{\sss BA}]})_\rmA - \Sigma_{\rmH }^{\perp \sss AB} \Big]
	+ \nu\cdot \lb \underline{\N}_{\sss M} (J^{\sss M})_\rmA - \rmJ_{\rmH }^\perp \rb \geq 0,
}
known as \emph{adiabaticity equation} \cite{Haehl:2014zda}. Directly putting the constitutive relations into this expression we can get,
\bee{
	\N_{\sss M} (S^{\sss M})_{\rmA} = 0.
}
Hence it suffices to choose an identically zero anomaly induced entropy current $(S^{\sss M})_{\rmA} = 0$, to satisfy the adiabaticity equation. 
We would like to comment here that vanishing of anomaly induced entropy current does not rely on
background being null; it is equally true for usual Einstein-Cartan backgrounds as well. See
\cref{belinEC} for more comments on the relativistic entropy current.

\paragraph*{Equilibrium Partition Function:}

In the beginning of this section we argued that at equilibrium, fluid can be described by a
partition function written purely in terms of background data. We would now attempt to find such an equilibrium partition function. We start by computing the variation of boundary effective action $S_{\cM,\text{anom}}$ given in \cref{S_anom_bdy},
\bem{
	\d S_{\cM,\text{anom}} = 
	\int \lbr \df x^{\sss M} \rbr \sqrt{|G|} \bigg[
		(T^{\sss MA})_\rmA \d E_{\sss AM}
		+ \lbr 
			(\Sigma^{\sss MAB})_{\rmA}
			- \Sigma_{\rmB\rmZ}^{\sss MAB}
		\rbr \d C_{\sss BMA}
		+ \lbr
			(J^{\sss M})_\rmA
			- \rmJ_{\rmB\rmZ}^{\sss M}
		\rbr \cdot \d A_{\sss M}
	\bigg] \\
	+ \int \lbr \df x^M \rbr \sqrt{|G|}
	\bigg[
		\hat \Sigma_{\rmB\rmZ }^{\sss MAB} \d \hat C_{\sss BMA}
		+ \hat\rmJ_{\rmB\rmZ }^{\sss M} \cdot \d \hat A_{\sss M}
	\bigg].
}
In equilibrium choosing transverse gauge for $\p_K$, i.e. $\L_{(K)} = [\L_{\Sigma(K)}]^{\sss A}_{\ \sss B} = 0$, the terms in last line vanish. Hence we can define the equilibrium boundary partition function,
\bee{\label{anom_eqbPF}
	W^{eqb}_{\cM,\text{anom}} = S_{\cM,\text{anom}}\bigg\vert_{\p_U = \p_K} = - \int_{\cM_{(2n+1)}} \frac{\bV}{\bH} \wedge \lb \bI^{(2n+2)} - \bm{{\hat I}}^{(2n+2)} \rb \bigg\vert_{\p_U = \p_K}.
}
Putting it together with $W_{\cB}$, we can get the equilibrium partition function for the full theory. In practice however, if one knows the expressions for Bardeen-Zumino currents, it suffices to have the boundary partition function to generate the covariant currents.

\subsection{Null Reduction -- Anomalous Galilean Fluids} \label{NC_fluids}

Having obtained the constitutive relations for anomalous null fluids, it is now time to perform null
reduction and extract out the Galilean results. To see this we can directly breakup the anomaly induced constitutive relations $(T^{\sss MA})_{\rmA}$, $(\Sigma^{\sss MAB})_\rmA$, $ (J^{\sss M})_\rmA$ into the basis given in \cref{redn_constitutive}. A straight away computation will yield trivial identifications,
\bee{\nn
	(\r^\mu)_{\rmA} = 0, \qquad
	(p^{\mu a})_{\rmA} = 0, \qquad
	(\t^{\mu a})_{\rmA} = 0, \qquad
	(s^\mu)_{\rmA} = 0, \qquad
}
\bee{\label{anom_consti}
	(\e^\mu)_{\rmA} = E^\mu_{\bcP}, \qquad
	(\s^{\mu ab})_{\rmA} = \Sigma^{\mu ab}_{\bcP}, \qquad
	(j^\mu)_{\rmA} =  J_{\bcP}^\mu.
}
We have also included an entropy current $(s^\mu)_{\rmA} = (S^\mu)_{\rmA}$ here which of-course is trivially zero. For the record we would write down the off-shell second law of thermodynamics for Galilean fluids,
\bem{\label{offshell_2ndLaw_gal}
	\underline{\Ndot}_\mu s^\mu
	+ \nu_{\sss\rmM} \underline{\Ndot}_\mu \r^\mu
	- \frac{1}{\vq} \lb 
		\underline{\Ndot}_\mu \e^{\mu}
		- [\text{power}]
		+ p^{\mu a} c_{\mu a}
	\rb
	+ \frac{1}{\vq} u^a \lb 
		\underline{\Ndot}_\mu p^{\mu}_{\ a} 
		- [\text{force}]_a
		+ \r^{\mu} c_{\mu a}
	\rb \\
	+ [\nu_{\t}]_{a} \lb \underline{\Ndot}_\mu \t^{\mu a} 
		- \half \lb \r^a - p^a \rb 
	\rb
	+ [\nu_\s]_{ba} \lb 
		\underline{\Ndot}_\mu \s^{\mu ab}
		- p^{[ba]}
		- 2 \t^{\mu[a} c_{\mu}^{\ b]}
		- \upsigma_{\rmH }^{\perp ab}
	\rb \\
	+ \nu \cdot \lb \underline{\Ndot}_\mu j^\mu - \mathrm j_{\rmH }^\perp \rb \geq 0,
}
where $u^\mu = \bar V_{(U)}^\mu$ (defined in \cref{defn_barV}) and $u^a = e^a_{\ \mu}
u^\mu$ is the spatial velocity of the fluid. $\vq$ is the temperature, $\nu_{\sss\rmM} = \vp -
\frac{1}{2\vq} u^a u_a$ is the total
mass chemical potential, $\vp$ is the mass chemical potential, $[\nu_\t]_a = [\nu_{\sss\Sigma}]^-_{\
a}$ is the boost chemical potential, $[\nu_\s]^a_{\ b} = [\nu_{\sss\Sigma}]^a_{\ b}$ is the spatial
spin chemical potential, and $\nu$ is the gauge chemical potential associated with fluid data $\p_U$ (find respective definitions in \cref{scaledCP,def_temperature}). This expression will hugely simplify if we choose $\p_T = \p_U$, i.e. choose to describe fluid in its local rest frame, because then $u^a = 0$,
\bem{\label{offshell_2ndLaw_lrf}
	\underline{\Ndot}_\mu s^\mu
	+ \vp \underline{\Ndot}_\mu \r^\mu
	- \frac{1}{\vq} \lb 
		\underline{\Ndot}_\mu \e^{\mu}
		- [\text{power}]
		+ p^{\mu a} c_{\mu a}
	\rb
	+ [\nu_{\t}]_{a} \lb \underline{\Ndot}_\mu \t^{\mu a} 
		- \half \lb \r^a - p^a \rb 
	\rb \\
	+ [\nu_\s]_{ba} \lb 
		\underline{\Ndot}_\mu \s^{\mu ab}
		- p^{[ba]}
		- 2 \t^{\mu[a} c_{\mu}^{\ b]}
		- \s_{\rmH }^{\perp ab}
	\rb
	+ \nu \cdot \lb \underline{\Ndot}_\mu j^\mu - j_{\rmH }^\perp \rb \geq 0.
}
It should be apparent that on putting in equations of motion it simply gives the second law of thermodynamics, $\underline{\Ndot}_\mu s^\mu \geq 0$. If one does not prefer to do reduction to get $(\e^\mu)_{\rmA}$, $(j^\mu)_{\rmA}, (\s^{\mu ab})_{\rmA}$, these can be generated directly from the Newton-Cartan transgression form,
\bee{
	\bcV_{\fp}^{(2n+1)} = - \frac{\bm n}{\bH} \wedge \lb \bm\fp^{(2n+2)} - \bm{{\hat\fp}}{}^{(2n+2)} \rb,
}
where $\bm\fp^{(2n+2)}$ is the NC anomaly polynomial defined at the end of \cref{null_inflow}, and hatted fields are,
\bee{
	\bm{{\hat A}} = \bA - \mu \bm n, \qquad
	\bm{{\hat C}}{}^a_{\ b} = \bC^a_{\ b} - [\mu_{\sss\s}]^a_{\ b} \bm n.
}
In terms of these anomaly induced constitutive relations can be generated as,
\bee{\label{anom_currents_NC}
	(j^\mu)_{\rmA}
	= *_\uparrow \lB \frac{\dow \bcV^{(2n+1)}_{\bm\fp}}{\dow \bF} \rB^\mu, \quad
	(\s^{\mu ab})_{\rmA}
	= *_\uparrow \lB \frac{\dow \bcV^{(2n+1)}_{\bm\fp}}{\dow \bR_{ba}}\rB^\mu, \quad
	(\e^\mu)_\rmA
	= *_\uparrow \lB \frac{\dow \bcV^{(2n+1)}_{\bm\fp}}{\dow \bH}\rB^\mu.
}
To write the equilibrium partition function in Newton-Cartan language we can use the natural
time-data in equilibrium $\p_T = \p_K = \p_U$. Hence using \cref{anom_eqbPF} we can find,
\bee{
	W^{eqb}_{\text{anom}} = - \int_{\cM^{K}_{(2n)}} \frac{\bm n}{\bH} \wedge \lb \bm \fii^{(2n+1)} - \bm{{\hat \fii}}^{(2n+1)} \rb \bigg\vert_{\p_U = \p_K},
}
where $\df \bm\fii^{(2n+1)} = \bm\fp^{(2n+2)}$; $\bm\fii^{(2n+1)}$ is just $\bI^{(2n+1)}_{\rmC\rmS
}$ after reduction. Please refer \cref{forms} for conventions on reducing the integral.

This concludes the main abstract results of this work. We have been able to construct gauge and gravitational anomalies in Galilean theories, and find their effect on Galilean hydrodynamics. We explicitly constructed the sector of fluid constitutive relations that is totally determined in terms of anomalies. These constitutive relations obey second law of thermodynamics with a trivially zero entropy current. We also found the equilibrium partition function which generates these constitutive relations in equilibrium configuration.

\section{Examples} \label{examples}

The entire discussion of this work till now has been very abstract. We will now try to illustrate it with few examples. In the following we will only discuss the case of abelian gauge field for simplicity. In \cref{walkthrough1} we start with a thorough walkthrough example for $3$ dimensional null theories (2 dimensional Galilean theories), where we perform each and every step as was done in the main work. We hope it will help the reader to understand the procedure more transparently. Later in \cref{example_gen} we present the results for arbitrary dimensional case up to next to leading order in derivatives.

\subsection{Walkthrough -- 2 Spatial Dimensions} \label{walkthrough1}

Let us go step by step for the case of 3 dimensional null backgrounds. The corresponding $5$ dimensional anomaly polynomial contains squared $\bF$ and $\bR$,
\bee{\label{anomP_5}
	\bcP^{(5)} 
	= \bm{{\bar V}} \wedge \lb 
		C^{(2)} \bF^{\wedge 2}
		+ C^{(2)}_g \bR^{\bar A}_{\ \bar B} \wedge \bR^{\bsss B}_{\ \bsss A}
	\rb,
}
from where we can read out the expression for $\bI^{(4)}$,
\bee{
	\bI^{(4)} 
	= - \bm{{\bar V}} \wedge \lB
		C^{(2)} \bA \wedge \bF
		+ C^{(2)}_g \lb
			\bm C^{\bsss A}_{\ \bsss B} \wedge \bR^{\bsss B}_{\ \bsss A}
			- \frac{1}{3} \bm C^{\bsss A}_{\ \bsss B} \wedge \bm C^{\bsss B}_{\ \bsss C} \wedge \bm C^{\bsss C}_{\ \bsss A}
		\rb
	\rB.
}
From here we can define the bulk partition function $W_{\cB} = \int_{ \cB_{(4)}} \bI^{(4)}$, and compute its variation (see \cref{Wbulk}),
\bem{
	\d W_{\cB}
	= \int_{ \cB_{(4)}} 2\lb 
		C^{(2)} \d\bA \wedge \bm{{\bar V}} \wedge \bF
		+ C^{(2)}_g \d\bm C^{\bsss A}_{\ \bsss B} \wedge \bm{{\bar V}}  \wedge \bR^{\bsss B}_{\ \bsss A}
	\rb \\
	- \int_{\cM_{(3)}} \lb 
		C^{(2)} \d\bA \wedge \bm{{\bar V}} \wedge  \bA
		+ C^{(2)}_g \d\bm C^{\sss A}_{\ \sss B} \wedge \bm{{\bar V}} \wedge  \bm C^{\sss B}_{\ \sss A}
	\rb.
}
Now using \cref{W_bulk} or \cref{hallcurrent_def}, we can find the Hall and Bardeen-Zumino currents,
\bea{
	\star_{(4)}\bm\rmJ_{\rmH } = \frac{\dow \bcP^{(5)}}{\dow \bF}
	= 2 C^{(2)} \bm{{\bar V}} \wedge \bF
	&\quad\Ra\quad \rmJ_{\rmH }^{\bar M} = C^{(2)} \e^{\bar N\bar R\bar S\bar M} \bar V_{\bar N}  F_{\bar R\bar S}, \nn\\
	\star_{(4)}\mathbf{\Sigma}_{\rmH }^{\bsss A \bsss B} = \frac{\dow \bcP^{(5)}}{\dow \bR_{\bsss B \bsss A}}
	= 2 C_g^{(2)} \bm{{\bar V}} \wedge \bR^{\bsss A \bsss B}
	&\quad\Ra\quad \Sigma_{\rmH }^{\bsss M \bsss A \bsss B} = C^{(2)}_g \e^{\bsss N\bsss R\bsss S\bsss M} \bar V_{\bsss N}  R_{\bsss R\bsss S}{}^{\bsss A \bsss B}, \nn\\
	\star\bm\rmJ_{\rmB\rmZ } = \frac{\dow \bI^{(4)}}{\dow \bF}
	= - C^{(2)} \bm{{\bar V}} \wedge \bA
	&\quad\Ra\quad \rmJ_{\rmB\rmZ }^{\sss M} = C^{(2)} \e^{\sss NRM} \bar V_{\sss N}  A_{\sss R}, \nn\\
	\star\mathbf{\Sigma}_{\rmB\rmZ }^{\sss AB} = \frac{\dow \bI^{(4)}}{\dow \bR_{\sss BA}}
	= - C_g^{(2)} \bm{{\bar V}} \wedge \bm C^{\sss AB}
	&\quad\Ra\quad \Sigma_{\rmB\rmZ }^{\sss MAB} = C^{(2)}_g \e^{\sss NRM} \bar V_{\sss N}  C^{\sss A \ B}_{\sss \hspace{0.2cm} R}.
}
The anomalous sources in \cref{EOM} are hence given as,
\bee{
	\Sigma^{\perp\sss AB}_{\rmH } = - C^{(2)}_g \e^{\sss MRS} \bar V_{\sss M}  R_{\sss RS}{}^{\sss AB}, \qquad
	\rmJ^\perp_{\rmH } = - C^{(2)} \e^{\sss MNR} \bar V_{\sss M}  F_{\sss NR}.
}
Here we have defined the volume element of the boundary manifold as $\e^{\perp \sss MNR} = \e^{\sss MNR}$. After null reduction we can trivially read out anomalous sources for NC conservation laws \cref{EOM_anom_NC},
\bee{
	\upsigma^{\perp ab}_{\rmH } = - C^{(2)}_g \ve^{\mu\nu}_\uparrow  R_{\mu\nu}{}^{ab}, \qquad
	\mathrm j^\perp_{\rmH } = - 2C^{(2)} \ve^{\mu\nu}_\uparrow  F_{\mu\nu}.
}

\paragraph*{Hydrodynamics:} We want to generate fluid constitutive relations which are compatible with anomalies described above. As described in the main text, it can be done using a transgression form  \cref{transgression},
\bea{
	\bcV^{(4)}_{\bcP}
	&= \frac{\bV}{\bH} \wedge \lb \bcP^{(5)} - \bm{{\hat\cP}}^{(5)} \rb \nn\\
	&= 
	- \bV \wedge \bm{{\bar V}} \wedge \lB
		2C^{(2)} \mu \bF
		+ 2C^{(2)}_g [\mu_{\sss\Sigma}]^{\bsss A}_{\ \bsss B}\bR^{\bsss B}_{\ \bsss A}
		+ \lb
			C^{(2)} \mu^2
			+ C^{(2)}_g [\mu_{\sss\Sigma}]^{\bsss A}_{\ \bsss B} [\mu_{\sss\Sigma}]^{\bsss B}_{\ \bsss A}
		\rb \bH
	\rB.
}
From its derivatives we can find various currents defined in \cref{anom_currents},
\bea{
	\star\bJ_{\bcP} = - 2C^{(2)} \mu \bV \wedge \bm{{\bar V}}
	&\quad\Ra\quad  J_{\bcP}^{\sss M} = 2C^{(2)} \mu \e^{\sss RSM}V_{\sss R} \bar V_{\sss S}, \nn\\
	\star\bm\Sigma_{\bcP}^{\sss AB} = - 2C^{(2)}_g [\mu_{\sss\Sigma}]^{\sss AB} \bV \wedge \bm{{\bar V}}
	&\quad\Ra\quad \Sigma_{\bcP}^{\sss MAB} = 2C^{(2)}_g  [\mu_{\sss\Sigma}]^{\sss AB} \e^{\sss RSM}V_{\sss R} \bar V_{\sss S}, \nn\\
	\star\bE_{\bcP} = - \lb
		C^{(2)} \mu^2
		+ C^{(2)}_g [\mu_{\sss\Sigma}]^{\sss A}_{\ \sss  B} [\mu_{\sss\Sigma}]^{\sss B}_{\ \sss A}
	\rb \bV \wedge \bm{{\bar V}}
	&\quad\Ra\quad E^{\sss M}_{\bcP} = \lb
		C^{(2)} \mu^2
		+ C^{(2)}_g [\mu_{\sss\Sigma}]^{\sss A}_{\ \sss  B} [\mu_{\sss\Sigma}]^{\sss B}_{\ \sss A}
	\rb \e^{\sss RSM}V_{\sss R} \bar V_{\sss S}.
}
Using \cref{anomT} we can trivially get the anomalous sector of constitutive relations from here.
These constitutive relations satisfy the adiabaticity equation \cref{adiabaticity} with zero entropy current, and at equilibrium also satisfy the anomalous equations of motion \cref{EOM}. Upon null reduction we can get the anomalous contribution to Galilean constitutive relations from here; the only surviving quantities are,
\bee{\nn
	(\e^\mu)_\rmA = \lb
		C^{(2)} \mu^2
		+ C^{(2)}_g [\mu_{\s}]^{a}_{\ b} [\mu_{\s}]^{b}_{\ a}
	\rb \ve^{\nu\mu}_\uparrow  n_\nu,
}
\bee{
	(\s^{\mu ab})_\rmA = 2C^{(2)}_g [\mu_{\s}]^{ab} \ve^{\nu\mu}_\uparrow  n_\nu, \qquad
	(j^\mu)_\rmA = 2C^{(2)} \mu \ve^{\nu\mu}_\uparrow n_\nu.
}
Finally we can write an equilibrium partition function $W^{eqb}_{\text{anom}}$ which generates these currents in equilibrium configuration. Using \cref{anom_eqbPF} we can directly find,
\bea{
	W^{eqb}_{\text{anom}} &= 
	- \int_{\cM_{(3)}} \frac{\bV}{\bH} \wedge \lb \bI^{(4)} - \bm{{\hat I}}^{(4)} \rb, \nn\\
	&= - \int_{\cM_{(3)}} \bV \wedge  \bm{{\bar V}} \wedge \lb 
		C^{(2)} \mu \bA
		+ C^{(2)}_g [\mu_{\sss\Sigma}]^{\sss A}_{\ \sss  B} \bm C^{\sss B}_{\sss \ A}
	\rb, \nn\\
	&= \int \df^3 x \sqrt{|\rmG|} \ \e^{\sss MNR} V_{\sss M} \bar V_{\sss N} \lb 
		C^{(2)} \mu  A_{\sss R}
		+ C^{(2)}_g [\mu_{\sss\Sigma}]^{\sss A}_{\sss \ B} C^{\sss B}_{\sss \ RA}
	\rb.
}
Same can be written in NC language,
\bea{
	W^{eqb}_{\text{anom}} 
	&= \int_{\cM_{(2)}} \bm n \wedge \lb 
		C^{(2)} \mu \bA
		+ C^{(2)}_g [\mu_{\s}]^a_{\ b} \bm C^b_{\ a}
	\rb, \nn\\
	&= \int \df^3 x \sqrt{|\g|} \ \ve^{\mu\nu}_\uparrow  n_\mu \lb 
		C^{(2)} \mu  A_\nu
		+ C^{(2)}_g [\mu_{\s}]^a_{\ b}  C^b_{\ \nu a}
	\rb.
}

\subsection{Arbitrary Even Spatial Dimensions up to Subsubleading Order} \label{example_gen}

Before proceeding with this example we should clarify the usage of `subsubleading' or `second
non-trivial' derivative order for null/Galilean fluids derived from relativistic fluids in
\cite{Banerjee:2015vxa}. One can check that in partition function or constitutive relations of a
$(2n+1)$-dim null fluid, first non-trivial contribution from parity odd-sector comes at $(n-1)$
derivatives, which is generally known as `leading parity-odd derivative order'. Correspondingly $n$
derivatives are called subleading while $(n+1)$ derivatives are called subsubleading. It is also
trivial to check that anomaly polynomial always has two more derivatives that the partition function
or constitutive relations. In anomalous sector one can check that first non-trivial contribution
comes at leading order (gauge anomaly) while no contribution comes at subleading order. Hence
`second non-trivial correction' comes at subsubleading order.

Coming back, one can check that up to subsubleading order $\bcP^{(2n+3)}$ and $\bI^{(2n+2)}$ (for $n>1$) are given as,
\bea{
	\bcP^{(2n+3)} 
	&= \bm{{\bar V}} \wedge \lb 
		C^{(2n)} \bF^{\wedge (n+1)}
		+ C^{(2n)}_g \bF^{\wedge (n-1)} \wedge \bR^{\bsss A}_{\ \bsss B} \wedge \bR^{\bsss B}_{\ \bsss A}
	\rb, \nn\\
	\bI^{(2n+2)} 
	&= - \bm{{\bar V}} \wedge \bA \wedge  \lb 
		C^{(2n)} \bF^{\wedge n}
		+ C^{(2n)}_g \bF^{\wedge (n-2)} \wedge \bR^{\bsss A}_{\ \bsss B} \wedge \bR^{\bsss B}_{\ \bsss A}
	\rb.
}
It would be worth noting that contribution from anomalies terminate at subsubleading order in $3$ spatial dimensions ($d=3,n=2$), hence these expressions are exact for $n=2$. From here we can get the Hall currents,
\bea{
	\Sigma_{\rmH }^{\perp \sss AB} &= - 2 C_g^{(2n)} \star\lB \bm{{\bar V}} \wedge \bF^{\wedge (n-1)} \wedge \bR^{\sss AB} \rB, \nn\\
	J^\perp_{\rmH } &= 
	- (n+1)C^{(2n)}  \star \lB
		\bm{{\bar V}} \wedge\bF^{\wedge n}
	\rB
	- (n-1) C^{(2n)}_g  \star \lB
		\bm{{\bar V}} \wedge\bF^{\wedge (n-2)} \wedge \bR^{\sss A}_{\ \sss B} \wedge \bR^{\sss B}_{\ \sss A}
	\rB,
}
that provide anomalies in \cref{EOM}. The results can be trivially transformed to Newton-Cartan language,
\bea{
	\upsigma^{\perp ab}_{\rmH } &= - 2 C_g^{(2n)} \ast_\uparrow\lB \bF^{\wedge (n-1)} \wedge \bR^{ab} \rB, \nn\\
	\mathrm j^\perp_{\rmH } &= 
	- (n+1)C^{(2n)}  \ast_\uparrow \lB
		\bF^{\wedge n}
	\rB
	- (n-1) C^{(2n)}_g  \ast_\uparrow\lB
		\bF^{\wedge (n-2)} \wedge \bR^a_{\ b} \wedge \bR^b_{\ a}
	\rB,
}
which provide anomalies in \cref{EOM_anom_NC}.

\paragraph*{Hydrodynamics:}
Using the anomaly polynomial one can find the constitutive relations for  the anomalous sector of hydrodynamics,
\bea{
	 J_{\bcP}^{\sss M}
	&=   
	(n+1)C^{(2n)} \sum_{m=1}^{n} {^n\bC_m} \mu^{m} \star\lB \bV \wedge \bm{{\bar V}} \wedge \bF^{\wedge (n-m)} \wedge \bH^{\wedge (m-1)} \rB^{\sss M} \nn\\ 
	&\quad
	+ (n-1) C^{(2n)}_g \lbr
		\sum_{m=1}^{n-2} {^{n-2}\bC_m} \mu^{m} \star\lB \bV \wedge \bm{{\bar V}} \wedge \bF^{\wedge (n-2-m)} \wedge \bR^{\sss A}_{\sss \ B} \wedge \bR^{\sss B}_{\sss \ A} \wedge \bH^{\wedge (m-1)} \rB^{\sss M}  \right. \nn\\ &\quad\left.
		+ \sum_{m=0}^{n-2} {^{n-2}\bC_m} \mu^{m}[\mu_{\sss\Sigma}]^A_{\ B} \star\lB \bV \wedge \bm{{\bar V}} \wedge \bF^{\wedge (n-2-m)} \wedge \lb 
			2 \bR^{\sss B}_{\sss \ A}
			+ [\mu_{\sss\Sigma}]^{\sss B}_{\sss \ A} \bH
		\rb \wedge \bH^{\wedge m} \rB^{\sss M}
	\rbr, \nn\\
	\Sigma_{\bcP}^{\sss MAB}
	&= 2 C_g^{(2n)} \lbr
		\sum_{m=1}^{n-1} {^{n-1}\bC_m} \mu^{m}  \star\lB \bV \wedge \bm{{\bar V}} \wedge \bF^{\wedge (n-1-m)} \wedge \bR^{\sss AB} \wedge \bH^{\wedge (m-1)} \rB^{\sss M}  \right. \nn\\ &\quad\left.
		+ \sum_{m=0}^{n-1} {^{n-1}\bC_m} \mu^{m} [\mu_{\sss\Sigma}]^{\sss AB} \star\lB \bV \wedge \bm{{\bar V}} \wedge \bF^{\wedge (n-1-m)} \wedge \bH^{\wedge m} \rB^{\sss M}
	\rbr, \nn\\
	E^{\sss M}_{\bcP}
	&= 
	\sum_{m=0}^{n-1} \mu^{m} \lb 
		{^{n+1}\bC_{m+2}}  C^{(2n)} \mu^2 
		+  {^{n-1}\bC_{m}}  C^{(2n)}_g  [\mu_{\sss\Sigma}]^{\sss A}_{\sss \ B} [\mu_{\sss\Sigma}]^{\sss B}_{\sss \ A}
	\rb \star\lB \bV \wedge \bm{{\bar V}} \wedge \bF^{\wedge (n-1-m)} \wedge \bH^{\wedge m} \rB^{\sss M} \nn\\
	&\quad
	+ C^{(2n)}_g \lbr 
		\sum_{m=2}^{n-1} {^{n-1}\bC_{m}} \mu^m \star\lB \bV \wedge \bm{{\bar V}} \wedge \bF^{\wedge (n-1-m)} \wedge \bR^{\sss A}_{\sss \ B} \wedge \bR^{\sss B}_{\sss \ A} \wedge \bH^{\wedge (m-2)} \rB^{\sss M} \right. \nn\\ &\quad\left.
		+ 2 \sum_{m=1}^{n-1} {^{n-1}\bC_{m}} \mu^m [\mu_{\sss\Sigma}]^{\sss A}_{\ \sss B} \star\lB \bV \wedge \bm{{\bar V}} \wedge \bF^{\wedge (n-1-m)} \wedge \bR^{\sss B}_{\sss \ A} \wedge \bH^{\wedge (m-1)} \rB^{\sss M}
	\rbr.
}
Anomalous sector of constitutive relations in terms of these are given by \cref{anomT}, while the entropy current is zero. Again by a trivial choice of basis these results can be transformed to Newton-Cartan basis; only non-zero constitutive relations are,
\bea{
	(j^\mu)_\rmA
	&=   
	(n+1)C^{(2n)} \sum_{m=1}^{n} {^n\bC_m} \mu^{m} \ast_\uparrow\lB \bm n \wedge \bF^{\wedge (n-m)} \wedge \bH^{\wedge (m-1)} \rB^\mu \nn\\ 
	&\quad
	+ (n-1) C^{(2n)}_g \lbr
		\sum_{m=1}^{n-2} {^{n-2}\bC_m} \mu^{m} \ast_\uparrow\lB \bm n \wedge \bF^{\wedge (n-2-m)} \wedge \bR^a_{\ b} \wedge \bR^b_{\ a} \wedge \bH^{\wedge (m-1)} \rB^\mu  \right. \nn\\ &\quad\left.
		+ \sum_{m=0}^{n-2} {^{n-2}\bC_m} \mu^{m} [\mu_{\s}]^a_{\ b} \ast_\uparrow\lB \bm n \wedge \bF^{\wedge (n-2-m)} \wedge \lb 
			2 \bR^b_{\ a}
			+ [\mu_{\s}]^b_{\ a} \bH
		\rb \wedge \bH^{\wedge m} \rB^\mu
	\rbr, \nn\\
	(\s^{\mu ab})_\rmA
	&= 2 C_g^{(2n)} \lbr
		\sum_{m=1}^{n-1} {^{n-1}\bC_m} \mu^{m}  \ast_\uparrow\lB \bm n \wedge \bF^{\wedge (n-1-m)} \wedge \bR^{ab} \wedge \bH^{\wedge (m-1)} \rB^\mu  \right. \nn\\ &\quad\left.
		+ \sum_{m=0}^{n-1} {^{n-1}\bC_m} \mu^{m} [\mu_{\s}]^{ab} \ast_\uparrow\lB \bm n \wedge \bF^{\wedge (n-1-m)} \wedge \bH^{\wedge m} \rB^\mu
	\rbr, \nn\\
	(\e^\mu)_{\rmA}
	&= 
	\sum_{m=0}^{n-1} \mu^{m} \lb 
		{^{n+1}\bC_{m+2}}  C^{(2n)} \mu^2 
		+  {^{n-1}\bC_{m}}  C^{(2n)}_g  [\mu_{\s}]^a_{\ b} [\mu_{\s}]^b_{\ a}
	\rb \ast_\uparrow\lB \bm n \wedge \bF^{\wedge (n-1-m)} \wedge \bH^{\wedge m} \rB^\mu \nn\\
	&\quad
	+ C^{(2n)}_g \lbr 
		\sum_{m=2}^{n-1} {^{n-1}\bC_{m}} \mu^m \ast_\uparrow\lB \bm n \wedge \bF^{\wedge (n-1-m)} \wedge \bR^a_{\ b} \wedge \bR^b_{\ a} \wedge \bH^{\wedge (m-2)} \rB^\mu \right. \nn\\ &\quad\left.
		+ 2 \sum_{m=1}^{n-1} {^{n-1}\bC_{m}} \mu^m [\mu_{\s}]^a_{\ b} \ast_\uparrow\lB \bm n \wedge \bF^{\wedge (n-1-m)} \wedge \bR^b_{\ a} \wedge \bH^{\wedge (m-1)} \rB^\mu
	\rbr.
}
Finally we can write an equilibrium partition function $W^{eqb}_{\text{anom}}$ which generates these currents in equilibrium configuration; for null fluids,
\bea{
	W_{\text{anom}}^{eqb} 
	&=  - \int_{\cM_{(2n+1)}} \bV \wedge \bm{{\bar V}} \wedge \bA \wedge  \lbr
		\sum_{m=1}^n {^{n}\bC_m} C^{(2n)} \mu^m \bm{{ F}}^{\wedge (n-m)} \wedge \bH^{\wedge (m-1)} \right. \nn\\ &\qquad \left.
		+ C^{(2n)}_g \sum_{m=1}^{n-2} {^{n-2}\bC_m} \mu^m \bF^{\wedge (n-2-m)} \wedge  \bH^{\wedge (m-1)} \wedge 
			\bR^{\sss A}_{\sss \ B} \wedge \bR^{\sss B}_{\sss \ A}  \right. \nn\\ &\qquad \left.
		+ C^{(2n)}_g \sum_{m=0}^{n-2} {^{n-2}\bC_m} \mu^m [\mu_{\sss\Sigma}]^{\sss A}_{\sss \ B}  \bF^{\wedge (n-2-m)} \wedge  \bH^{\wedge m} \wedge 
		\lb 
			2\bR^{\sss B}_{\sss \ A}
			+ [\mu_{\sss\Sigma}]^{\sss B}_{\sss \ A} \bH
		\rb
	\rbr,
}
and for Galilean fluids,
\bea{
	W_{\text{anom}}^{eqb} 
	&=  - \int_{\cM_{(2n)}} \bm n \wedge \bA \wedge  \lbr
		\sum_{m=1}^n {^{n}\bC_m} C^{(2n)} \mu^m \bm{{ F}}^{\wedge (n-m)} \wedge \bH^{\wedge (m-1)} \right. \nn\\ &\qquad \left.
		+ C^{(2n)}_g \sum_{m=1}^{n-2} {^{n-2}\bC_m} \mu^m \bF^{\wedge (n-2-m)} \wedge  \bH^{\wedge (m-1)} \wedge 
			\bR^{a}_{\ b} \wedge \bR^{b}_{\ a}  \right. \nn\\ &\qquad \left.
		+ C^{(2n)}_g \sum_{m=0}^{n-2} {^{n-2}\bC_m} \mu^m [\mu_{\s}]^{a}_{\ b}  \bF^{\wedge (n-2-m)} \wedge  \bH^{\wedge m} \wedge 
		\lb 
			2\bR^{b}_{\ a}
			+ [\mu_{\s}]^{b}_{\ a} \bH
		\rb
	\rbr.
}
This finishes our discussion about anomalies in generic even dimensional Galilean fluid up to subsubleading order in derivative expansion. $1$ spatial dimensional case was discussed separately in \cref{walkthrough1} for illustrative purposes. $1$ dimensional case is also qualitatively different from higher dimensions, because only in this special case we get pure gravitational anomaly term in the anomaly polynomial up to subsubleading order. $3$ and higher spatial dimensional cases are qualitatively similar as we illustrated above. For physically interesting results one might want to put $n=2$ and recover $3$ spatial dimensional results, which are found to be in agreement with path integral calculation of \cite{Bakas:2011nq}.

\section{Conclusions and Further Directions}

In this work we examined the effect of gauge and gravitational anomalies on Galilean theories with spin current, coupled to torsional Newton-Cartan geometries. In particular it is to be noted that we primarily studied anomalous theories on torsional null backgrounds, from where the aforementioned system is just a choice of basis (null reduction) away. It strengthens our belief that null theories are just an embedding of Galilean theories into a higher dimensional spacetime, which are closer to their relativistic cousins, are frame independent and are easier to handle compared to Newton-Cartan backgrounds. Transition from null to Galilean (Newton-Cartan) theories is essentially trivial.

We used the anomaly inflow mechanism prevalent in relativistic theories, with slight modifications, to construct these anomalies. We found that after null reduction the anomalies only contribute to spatial spin and charge conservation equations, and only in even dimensions. In other words only rotational and gauge symmetry of the Galilean theory goes anomalous. This is in contrast with the results of \cite{Jensen:2014hqa} where Galilean boost symmetry was also seen to be anomalous. As we mentioned in the introduction and in the main work, the discrepancy can be attributed to presence of extra fields in \cite{Jensen:2014hqa} which have been explicitly switched off in null background construction. It is interesting to note that Galilean anomaly polynomial $\bm\fp^{(2n+2)}$ is structurally the same as relativistic anomaly polynomial $\bcP_{\rmC\rmS }^{(2n+2)}$, and hence the number on anomaly coefficients at the both sides match. Owing to it, the structure of Hall currents that enter the conservation laws, is also quite similar in both cases. Hence the results we have obtained promises to be genuine non-relativistic anomalies and not just the manifestation of (stronger) Galilean invariance.

Unrelated to Galilean theories, we found that in Cartan formulation of relativistic fluids there
exists a more natural definition of entropy current which does not get any anomalous
contributions. On the other hand the Belinfante (usual) entropy current used e.g. in
\cite{Haehl:2015pja} gets contributions from gravitational anomaly. Look at \cref{belinEC} for more
comments on this issue.

We also studied anomalous sector of null/Galilean hydrodynamics, in which we explicitly wrote down the constitutive relations which are completely determined in terms of anomalies. For this we used the transgression machinery developed to do the same task in relativistic hydrodynamics. There have been no surprises in this computation; everything went more or less through for null theories, as it did for relativistic theories. The entropy current in Galilean theories is independent of anomalies as well.
From a different perspective, it illustrates that null background construction allows us to use much sophisticated and developed relativistic machinery directly into non-relativistic physics, which is encouraging.

It opens up an arena to bring in set results from relativistic theories into null theories and see
if we can say something new and useful about Galilean theories from there. An immediate question
that comes to our mind is regarding transcendental contributions to hydrodynamics from anomalies. In
relativistic hydrodynamics \cite{Jensen:2012kj} showed that there are certain constants in fluid
constitutive relations that are left undetermined by second law of thermodynamics, but can be
related to the anomaly coefficients on requiring the consistency of Euclidean vacuum. Similar
constants have also been found for Galilean theories in \cite{Banerjee:2015hra,Jensen:2014ama}. It
would be nice to see if these constants can be associated with Galilean anomalies found in this work. Being little more ambitious, one can hope for a complete classification of Galilean hydrodynamic transport following its relativistic counterpart suggested recently in \cite{Haehl:2015pja,Haehl:2014zda}. It will also be interesting to see if Weyl anomaly analysis of \cite{Jensen:2014hqa} remains unchanged when the additional mass sources have been switched off.

For now we will leave the reader with these questions and possibilities, in hope that we would be able to unravel new and interesting non-relativistic physics using null backgrounds. If there is one thing reader should take away from this work, we would recommend the approach -- if we are interested in a problem pertaining to Galilean physics which we know how to solve in relativistic case, a good way ahead would be to formulate the problem in terms of null theories, do the computation there, and perform a trivial null reduction to get Galilean results.

\subsection*{Acknowledgments}

Author wishes to thank Nabamita Banerjee and Suvankar Dutta for their unconditional guidance and
support, which made this work possible. Author also wishes to acknowledge many useful discussions
with Felix Haehl, Kristan Jensen and R. Loganayagam. AJ is financially supported by the Durham Doctoral Scholarship, and would like to acknowledge the hospitality of IISER Bhopal where part of this project was done.

\appendix

\section{Results in Non-Covariant Basis} \label{non_cov}

In this appendix we give some of the results discussed in main text in conventional non-covariant notation. We pick up  a basis $x^{\sss M} = \{x^{\sss\sim}, t,x^i \}$ on $\cM_{(d+2)}$ such that time data $\p_V = \{ \dow_{\sss\sim}, 0, 0 \}$ and $\p_T = \{\dow_t, 0 , 0\}$. $\vec x = \{ x^i \}$ spans the spatial slice $\cM_{(d)}^T$. This is equivalent to choosing the Newton-Cartan decomposition but with $v^i = \L_{(T)} = [\L_{\Sigma(T)}]^{\sss A}_{\ \sss B} = 0$. On $\bbR^{(d+1,1)}$ on the other hand we choose the same basis as before $x^{\sss A} = \{ x^-, x^+, x^a \}$ such that $V = \dow_-$, $\bar V_{(T)} = \dow_+$.
In this basis various NC background fields can be decomposed as,
\bee{\nn
	 n_\mu = \begin{pmatrix} \E{-\F} \\ \E{-\F} a_i \end{pmatrix}, \qquad
	 v^\mu = \begin{pmatrix} \E{\F} \\ 0 \end{pmatrix}, \qquad
	 B_\mu = \begin{pmatrix}  B_t \\  B_i \end{pmatrix}, \qquad
	e^a_{\ \mu} = \begin{pmatrix} 0 \\ e^a_{\ i} \end{pmatrix}, \qquad
	e_a^{\ \mu} = \begin{pmatrix} - a_a \\ e_a^{\ i} \end{pmatrix},
}
\bee{
	h_{\mu\nu} = \begin{pmatrix}
		0 & 0 \\
		0 & g_{ij}
	\end{pmatrix}, \qquad
	h^{\mu\nu} = \begin{pmatrix}
		a^k a_k & - a^j \\
		- a^i & g^{ij}
	\end{pmatrix}.
}
Here spatial metric has been defined as,
\bee{
	g_{ij} = \d_{ab}e^a_{\ i}e^b_{\ j}, \qquad
	g^{ij} = \d^{ab}e_a^{\ i}e_b^{\ j}.
}
Spatial indices can be raised and lowered by $g_{ij}$ and can be swapped using the spatial Vielbein $e^a_{\ i}$. However in the following we will explicitly work in $i,j\ldots$ indices. One can check that after this choice of basis we are only allowed to perform $\vec x$ dependent transformations, except boosts which are completely fixed. One can check that on trivially decomposing the Newton-Cartan expressions into $x^\mu = \{ t, x^i \}$, our theory will be manifestly covariant against all these transformations except time translations $t \ra t + \xi^t(\vec x)$, sometimes known as Kaluza-Klein (KK) gauge transformations. The theory can be made manifestly covariant under KK transformations as well by working with corrected tensors,
\bee{
	\acute{X}^t = \E{-\F} \lb X^t + a_i X^i \rb, \qquad
	\acute{X}_t = \E{\F} X_t, \qquad
	\acute{X}^i = X^i, \qquad
	\acute{X}_i = X_i - a_i X_t,
}
and similarly for higher rank tensors. These are the well known Kaluza-Klein covariant
fields\footnote{The original Kaluza-Klein transformation only involves KK gauge field $a_i$. The
  factors of $\E{\F}$ can be thought of as red-shift factors due to time component of time metric $n_\mu$.}. Under flat time approximation i.e. $\F = a_i = 0$ this transformation turns trivial. One can check that NC contraction can be expanded in this format as,
\bee{
	A^\mu B_\mu = \acute A^t \acute B_t + \acute A^i \acute B_i,
}
which will be helpful later.
Now we can decompose various components of connections in this basis as,
\bee{\nn
	c_{\mu t} = 0, \qquad
	\acute c_{ij}
	= 
	\half \acute\dow_t g_{ji}
	+ \half \acute\O_{ij}
	+ \acute\rmT_{(ij)t}, \qquad
	\acute\a_i = \acute c_{ti} = \acute\O_{ti},
}
\bee{\nn
	\acute\G^{t}_{\ tt} = - \acute \dow_{t} \F, \qquad
	\acute\G^{t}_{\ tj} = \E{-\F} \acute\dow_{t} a_{j}, \qquad
	\acute\G^{t}_{\ it} = - \acute \dow_{i} \F , \qquad
	\acute\G^{t}_{\ ij} = \E{-\F} \acute\dow_{i} a_{j}.
}
\bee{\nn
	\acute\G^{k}_{\ tt}
	= 
	\acute \O_{t}^{\ k}, \qquad
	\acute\G^{k}_{\ it}
	= 
	\acute c_{i}^{\ k}, \qquad
	\acute\G^{k}_{\ tj}
	= 
	\acute c_{j}^{\ k}
	- \acute\rmT_{\ jt}^{k}.
}
\bee{
	\acute\G^{k}_{\ ij}
	= 
	\half g^{kl} \lb \acute\dow_i g_{lj} + \acute\dow_j g_{li} - \acute\dow_l g_{ij} \rb
	+ \half \lb 
		\acute\rmT_{\ ij}^{k}
		- 2 \acute\rmT_{(ij)}^{\ \ \ k}
	\rb.
}
Here we have also defined the corrected coordinate derivatives $\acute\dow_i = \dow_i - a_i \dow_t$,
$\acute\dow_t = \E{\F}\dow_t$. In \emph{equilibrium} (i.e. $\dow_t\vf = 0 \ \forall \f$) or when time
is flat, we can recover $\acute\dow_i = \dow_i$, $\acute\dow_t = \dow_t$. We define covariant derivative $\acute\N_i$ associated with corrected derivative\footnote{
One might be lured (e.g. in \cite{Banerjee:2015hra}) to define covariant 
derivative with respect to original derivative $\dow_i$ and more conventional affine connection,
\bee{
	\g^{k}_{\ ij}
	= \acute\G^{k}_{\ ij}
	- \half a^k \dow_t g_{ij}
	+ g^{kl} a_{(i} \dow_t g_{j)l}
	= 
	\half g^{kl} \lb \dow_i g_{lj} + \dow_j g_{li} - \dow_l g_{ij} \rb
	+ \half \lb 
		\acute\rmT_{\ ij}^{k}
		- 2 \acute\rmT_{(ij)}^{\ \ \ k}
	\rb,
}
which however will not be KK gauge invariant. The results hence will be messy and will carry extra time derivatives of metric. Therefore we will refrain from doing so. Obviously both of these covariant derivatives are same in flat time or equilibrium.
} 
$\acute\dow_i$ and connections $\acute\G^j_{\ ik}$ and $\acute A_i$, which act on a general tensor $\vf^{i}_{\ j}$ transforming in adjoint representation of the gauge group,
\bee{
	\acute\N_i \tensor{\vf}{^j_k} = \acute\dow_i \tensor{\vf}{^j_k}
	+ \acute\G^j_{\ il}\tensor{\vf}{^l_k}
	- \acute\G^l_{\ ik}\tensor{\vf}{^j_l}
	+ \lB \acute A_i , \tensor{\vf}{^j_k} \rB,
}
and similarly on higher rank objects. 
We also define a `time covariant derivative' $\acute\N_t$ associated with $\acute\dow_t$ and
connections $\acute\G^j_{\ tk}$ and $ \acute A_t$, acting on $\vf^{i}_{\ j}$ naturally,
\bee{
	\acute\N_t \tensor{\vf}{^i_j} = \acute \dow_t \tensor{\vf}{^i_j}
	+ \acute\G^i_{\ tk}\tensor{\vf}{^k_j}
	- \acute\G^k_{\ tj}\tensor{\vf}{^i_k}
	+ \lB  \acute A_t , \tensor{\vf}{^i_j} \rB,
}
and similarly on higher rank objects. 
One can check that both of these derivatives behave tensorially on the spatial slice and are KK gauge invariant. More importantly both of these preserve metric $g_{ij}$. 
There is no essential need to work with these corrected quantities, just that the statements are manifestly KK gauge invariant and look nicer.

Using similar decomposition and KK correction for various currents, we can reduce the conservation
equations \cref{EOM_anom_NC} into non-covariant basis,
\bea{\label{EOM_anom_noncov}
	\text{Mass Cons.:}&\quad 
	\underline{\acute\N}_t \r
	+ \underline{\acute\N}_i \acute\r^i
	= 0, \nn\\
	\text{Energy Cons.:}&\quad 
	\underline{\acute\N}_t \e
	+ \underline{\acute\N}_i \acute\e^i
	=
	[\text{power}]
	- \acute p^{i} \acute\a_{i}
	- \acute p^{ij} \acute c_{ij}, \nn\\
	\text{Momentum Cons.:}&\quad 
	\underline{\acute\N}_t \acute p_{i}
	+ \underline{\acute\N}_j \acute p^{j}_{\ i}
	=
	[\text{force}]_i
	- \r \acute\a_{i}
	- \acute \r^{j} \acute c_{ji}, \nn\\
	\text{Temporal Spin Cons.:}&\quad 
	\underline{\acute\N}_t \acute\t^i
	+ \underline{\acute\N}_j \acute\t^{ji}
	= 
	\half \lb \acute \r^i - \acute p^i\rb, \nn\\
	\text{Spatial Spin Cons.:}&\quad  
	\underline{\acute\N}_t \acute\s^{ij}
	+ \underline{\acute\N}_k \acute\s^{kij}
	 = 
	\acute p^{[ji]}
	+ 2 \acute\t^{[i} \acute \a^{j]}
	+ 2 \acute\t^{k[i} \acute c_{k}^{\ j]}
	+ \acute\upsigma^{\perp ij}_{\rmH }, \nn\\
	\text{Charge Cons.:}&\quad  
	\underline{\acute\N}_t q
	+ \underline{\acute\N}_i \acute j^i = \mathrm j_{\rmH }^\perp,
}
where $\underline{\acute\N}_i = \acute\N_i - \acute\rmT^{j}_{\ ji} + \acute H_{ti}$ and
$\underline{\acute\N}_t = \acute\N_t + \acute\G^i_{\ ti}$. It is worth noting that the corrected
time component of mass current $\acute\r^t$ is just the mass density $\r$, and similarly for all
other currents. If we are to expand the covariant derivatives in these equations, the nice looking
expressions will turn notoriously bad; so we do not attempt to do that. Rather we invite the readers
to qualitatively access the form of these equations and convince themselves that these are what we
expect for a Galilean system. Similarly [power] and [force] densities can also be decomposed as,
\bee{\label{power.work_noncov}
	[\text{power}] = \acute e_i \cdot \acute j^i + \ldots, \qquad
	[\text{force}]^i = \acute e_i \cdot q + \acute \b_{ij}\cdot \acute j^j  + \ldots,
}
where $\acute e_i = \acute F_{it}$ is the electric field, $\acute\b_{ij} = \acute F_{ij}$ is the dual magnetic field, and $\ldots$ corresponds to similar terms coming from all other field-current couples.

On the other hand non-covariant expressions for anomalous sector of hydrodynamic constitutive relations follow trivially from \cref{anom_consti}. Only non-zero contributions are given as\footnote{We have assumed that the same $\p_T$ is being used for reduction and to describe anomaly polynomial. Had they been different, currents would shift by a total derivative.},
\bee{
	(\acute\e^i)_{\rmA} = E_{\bcP}^i, \qquad
	(\acute\s^{ijk})_{\rmA} = \Sigma_{\bcP}^{ijk}, \qquad
	(\acute j^i)_{\rmA} =  J_{\bcP}^i.
}
The expressions for RHS can be obtained from \cref{anom_currents_NC}.

\section{Comparison with Geracie et al. \cite{Geracie:2015xfa}} \label{Geracie:2015xfa}

Authors of \cite{Geracie:2015xfa} have prescribed a nice covariant frame independent description of Galilean physics in terms of an `extended representation'. The extended space is basically a one dimensional higher flat space which allows for a nice frame independent embedding of the Galilean group. On a closer inspection however, it would be clear that the extended space is nothing but the Vielbein space for null theories. To demonstrate this we pick up a basis on $\cM_{(d+2)}$ (but not perform null reduction, which would require us to choose time data $\p_T$ and hence will introduce frame dependence), $x^{\sss M} = \{ x^{\sss\sim}, x^\mu \}$ such that $V = \dow_{\sss \sim}$. We can then express the anomalous null conservation laws as,
\bea{
	\lb \Ndot_\mu - \rmT^\nu_{\ \nu\mu} \rb j^\mu_\r
	&=
	0, \nn\\
	\rmE^{\sss A}_{\ \nu}  \lb \Ndot_\mu - \rmT^\nu_{\ \nu\mu} \rb  T^{\mu}_{\ \sss A}
	- \rmT^{\sss A}_{\ \nu\mu}  T^{\mu}_{\ \sss A}
	&=
	 R_{\nu\mu}{}^{\sss A}_{\sss \ B} \Sigma^{\mu \sss  B}_{\sss \ \ \ A}
	+  F_{\nu\mu}\cdot  J^\mu, \nn\\
	\lb \Ndot_\mu - \rmT^\nu_{\ \nu\mu} \rb \Sigma^{\mu \sss  AB} &= -  T^{[{\sss AB}]} + \Sigma_{\rmH }^{\perp \sss  AB} + \#^{[\sss A} V^{{\sss B}]}, \nn\\
	\lb \Ndot_\mu - \rmT^\nu_{\ \nu\mu} \rb j_q^\mu &= \rmJ_{\rmH }^\perp.
}
In this, and only in this section $\Ndot_\mu$ is associated with $\G^\r_{\ \mu\s}$, $C^{\sss A}_{\ \mu \sss B}$, $A_\mu$ and Vielbein has been used to transform indices. The results are presented to make them look as close as possible to (5.8) -- (5.10) of \cite{Geracie:2015xfa}. The authors however did not consider anomalies, and did not report the full spin conservation. Only the boost part of spin conservation is reported in (5.13) of \cite{Geracie:2015xfa} which is identical to our corresponding conservation in \cref{EOM_anom_NC}.

If one looks at these equations and at the currents appearing in them, one would realize that all the unphysical degrees of freedom have been eliminated (except the spin conservation equations). Therefore EM tensor and charge current as appearing in \cite{Geracie:2015xfa} only carry physical information. On cost of some unphysical degrees of freedom (and a consistent prescription to eliminate them) we have been able to transform these set of equations into a nice covariant higher dimensional null theory.

We would like to note that authors in \cite{Geracie:2015xfa} have also used their construction to study $(2+1)$ dimensional Galilean fluids. Same results (for torsionless case) were gained from `null fluids' in \cite{Banerjee:2015hra} and a detailed comparison can be found in their last appendix.

\section{Conventions of Differential Forms} \label{forms}

In this appendix we will recollect some results about differential forms, and will set notations and conventions used throughout this work. An $m$-rank differential form $\bm\mu^{(m)}$ on $\cM_{(d+2)}$, can be written in a coordinate basis as,
\bee{
	\bm\mu^{(m)} = \frac{1}{m!}\mu_{\sss M_1M_2\ldots M_m} \df x^{M_1} \wedge \df x^{\sss M_2} \wedge \ldots \wedge \df x^{\sss M_m},
}
where $\mu$ is a completely antisymmetric tensor. On $\cM_{(d+2)}$, volume element is given by a full rank form, 
\bee{
	\bm\e^{(d+2)} = \frac{1}{(d+2)!}\e_{\sss M_1M_2\ldots M_{d+2}} \df x^{\sss M_1} \wedge \df x^{\sss M_2} \wedge \ldots \wedge \df x^{\sss M_{d+2}},
}
where $\e$ is the totally antisymmetric Levi-Civita symbol with value $\e_{0,1,2,\ldots,d+1} = \sqrt{|\rmG|}$ and $\rmG =\det{\rmG_{MN}}$. Using it,  Hodge dual is defined to be a map from $m$-rank differential forms to $(d+2-m)$-rank differential forms,
\bee{
	\star[\bm\mu^{(m)}] = \frac{1}{(d+2-m)!} \lb\frac{1}{m!}\mu^{\sss M_1\ldots M_m} \e_{\sss M_1\ldots M_{m}N_{1}\ldots N_{d+2-m}} \rb 
	\df x^{\sss N_{1}} \wedge \ldots \wedge \df x^{\sss N_{d+2-m}}.
}
One can check that $\star\star\bm\mu^{(m)} = \sgn(\rmG) (-)^{m(d-m)}$, and,
\bee{
	\bm\mu^{(m)} \wedge \star [\bm\nu^{(m)}] =  \frac{1}{m!} \mu^{\sss M_1\ldots M_m} \nu_{\sss M_1\ldots M_m} \bm\e^{(d+2)}.
}
We define the $\wedge$ product of two differential forms as,
\bee{
	\bm\mu^{(m)} \wedge \bm\nu^{(r)}
	= 
	\frac{1}{(m+r)!}\lb\frac{(m+n)!}{m!r!}\mu_{[\sss M_1\ldots M_m} \nu_{{\sss N_1\ldots N_r}]} \rb
	\df x^{\sss M_1} \wedge \ldots \wedge \df x^{\sss N_1} \wedge \ldots.
}
For multiple forms we can find,
\bem{
	\bm\mu^{(m)} \wedge \bm\nu^{(r)} \wedge \ldots \wedge \bm\r^{(s)}
	= 
	\frac{1}{(m+r+\ldots+s)!}\lb\frac{(m+r+\ldots+s)!}{m!r!\ldots s!}
	\mu_{[\sss M_1\ldots M_m} \nu_{\sss N_1\ldots N_r} \r_{{\sss R_1\ldots R_s}]} \rb \\
	\df x^{\sss M_1} \wedge \ldots \wedge \df x^{\sss N_1} \wedge \ldots \wedge \df x^{\sss R_1} \wedge \ldots,
}
\bem{
	\star[\bm\mu^{(m)} \wedge \bm\nu^{(r)} \ldots \wedge \bm\r^{(s)}]
	= \frac{1}{(d+2-m-r\ldots-s)!} 
	\lb\frac{1}{m!r!\ldots s!} \mu^{\sss M_1\ldots} \nu^{\sss N_1\ldots} \ldots \r^{\sss R_1\ldots} \e_{\sss M_1\ldots N_{1}\ldots R_1\ldots \cdots S_1\ldots} \rb \\
	\df x^{\sss S_{1}} \wedge \ldots.
}
We define the interior product with respect to a vector field $X$ of a differential form as,
\bee{
	\i_X \bm\mu^{(m)} = \frac{1}{(m-1)!}\lb X^{\sss M}\mu_{[{\sss MN_1\ldots N_{m-1}}]} \rb \df x^{\sss N_1}  \wedge \ldots \wedge \df x^{\sss N_{m-1}}.
}
One can check two useful identities
\bee{
	\i_X \star\lB\bm\mu^{(m)}\rB = \star\lB\bm\mu^{(m)} \wedge \bm X \rB, \qquad
	\star\lB \i_X \bm\mu^{(m)} \rB = (-)^{m-1} \bX \wedge \star \lB \bm\mu^{(m)} \rB.
}
Given a one-form $\bY^{(1)}$ and a vector field $X$ such that $\i_X \bY^{(1)} = 1$, any differential form $\bm\mu^{(m)}$ can be decomposed as,
\bee{
	\bm\mu^{(m)} = \i_X \lb \bY^{(1)} \wedge \bm\mu^{(m)} \rb + \bY^{(1)} \wedge \i_X \bm\mu^{(m)}.
}
This is in particular helpful when $\bm\mu^{(d+2)}$ is a full rank form,
\bee{
	\bm\mu^{(d+2)} =  \bY^{(1)} \wedge \i_X \bm\mu^{(d+2)}.
}
The exterior product of a differential form is defined to be,
\bee{
	\df \bm\mu^{(m)} = \frac{1}{(m+1)!} \lB (m+1) \dow_{[\sss M_1} \mu_{{\sss M_2\ldots M_{m+1}}]} \rB \df x^{\sss M_1} \wedge \ldots \wedge \df x^{\sss M_{p+1}}.
}
One can check a useful relation,
\bee{
	\star \df \bm\mu^{(d+1)} = (-)^{d+1} \underline{\N}_{\sss M} \star \lB\bm\mu^{(d+1)}\rB^{\sss M}, \qquad
	\df \star \lB\bm\mu^{(1)}\rB = \star \underline{\N}_{\sss M} \mu^{\sss M}.
}
The lie derivative of a differential form satisfies,
\bee{
	\lie_X \bm\mu^{(m)} = \i_X \df \bm\mu^{(m)} + \df \lb \i_X\bm\mu^{(m)} \rb.
}
Integration of a full rank form is defined as,
\bea{
	\int_{\cM_{(d+2)}} \bm\mu^{(d+2)} 
	&= \sgn(\rmG)\int \lbr \df x^{\sss M} \rbr \sqrt{|\rmG|} \ \star[\bm\mu^{(d+2)}]  \nn\\
	&= \sgn(\rmG)\int \lbr \df x^{\sss M} \rbr \sqrt{|\rmG|} \ \frac{1}{(d+2)!}\e^{\sss M_1\ldots M_{d+2}}\mu_{\sss M_1\ldots M_{d+2}}.
}
Here the raised Levi-Civita symbol has value $\e^{0,1,2,\ldots,d+1} = \sgn(G)/\sqrt{|G|}$. Integration of an exact full rank form is given by integration on the boundary,
\bee{
	\int_{\cM_{(d+2)}} \df\bm\mu^{(d+1)} 
	= \int_{\dow\cM_{(d+2)}} \bm\mu^{(d+1)},
}
where given a unit vector $N$ normal to boundary, volume element on the boundary is defined as $\i_N \bm\e^{(d+2)} = \star \bN$.

\subsection*{Newton-Cartan Differential Forms}

We decompose a vector and a one form on $\cM_{(d+2)}$ in NC basis,
\bea{
	\cX^M \dow_M &=  \lb \cX^\sim -  B_\mu \cX^\mu \rb \dow_\sim + \cX^\mu \lb \dow_\mu +  B_\mu \dow_\sim \rb, \nn\\
	\cY_M \df x^M &= \cY_\sim \lb \df x^\sim -  B_\mu \df x^\mu \rb + \lb \cY_\mu +  B_\mu \cY_\sim \rb  \df x^\mu.
}
One can check that these results are written in `nicely' transforming basis from the NC perspective, which tells us that,
\bee{
	V^M\cY_M = \cY_\sim, \qquad
	\cY_\mu +  B_\mu \cY_\sim, \qquad
	\bar V_M\cX^M = \cX^\sim - X^\mu  B_\mu, \qquad
	\cX^\mu,
}
are nicely transforming quantities. As is quite apparent, first and last do not depend on the explicit choice of $\p_T$ but the middle ones do. A similar analysis can be done for all tensor fields in the theory. Note that if $\cY_M$ satisfies $\i_V \bcY = V^M\cY_M = 0$, the one form becomes purely NC. On the other hand if $\bar V_M\cX^M = 0$, the vector field becomes purely NC. This motivates us to define a NC differential form to be a form in $\cM_{(d+2)}$ which does not have a leg along $V$, i.e. $\i_V \bm\mu^{(m)}$. Such a form can be expanded as,
\bee{
	\bm\mu^{(m)} = \frac{1}{m!}\mu_{\mu_1\mu_2\ldots \mu_m} \df x^{\mu_1} \wedge \df x^{\mu_2} \wedge \ldots \wedge \df x^{\mu_m}.
}
On the other hand we define a NC `differential contra-form' as a totally antisymmetric contravariant tensor in $\cM_{(d+2)}$ which has zero contraction with $\bar V_M$. In basis $\dow'_\mu = \dow_\mu +  B_\mu \dow_\sim$ it can be expanded as,
\bee{
	\bm\mu^{[m]} = \frac{1}{m!}\mu^{\mu_1\mu_2\ldots \mu_m} \dow'_{\mu_1} \wedge \dow'_{\mu_2} \wedge \ldots \wedge \dow'_{\mu_m}.
}
It is clear that though the basis depends on choice of $\p_T$, the components of contra-form are independent of it. On a manifold with a non-degenerate metric there exists a map between these two quantities, but for us these two shall be distinct. We can also define a spatial differential form/contra-form with requirement that it should not have any leg along $V$ and $\bar V$. In this case there exists a map between these two quantities realized by $p^{\mu\nu}$ and $p_{\mu\nu}$.

Correspondingly there are three volume elements,
\bea{
	\bm\ve^{[d+1]}_\uparrow &= [\star\bm{{\bar V}}]^\sharp = \frac{1}{(d+1)!} \lb \bar V_{M} \e^{M \mu_{1}\ldots \mu_{d+1}} \rb 
	\dow'_{\mu_1} \wedge \ldots \wedge \dow'_{\mu_{d+1}}, \nn\\
	\bm\ve^{(d+1)}_\downarrow &= \star\bV = \frac{1}{(d+1)!} \lb V^{M} \e_{M \mu_{1}\ldots \mu_{d+1}} \rb 
	\df x^{\mu_{1}} \wedge \ldots \wedge \df x^{\mu_{d+1}}, \nn\\
	\bm \ve^{(d)} &= \star[\bV \wedge \bm{{\bar V}}] = \frac{1}{d!} \lb V^{M} \bar V^N \e_{MN \mu_{1}\ldots \mu_{d}} \rb 
	\df x^{\mu_{1}} \wedge \ldots \wedge \df x^{\mu_{d}}.
}
In the main text we have primarily used the first one. Correspondingly there are three Hodge duals that provide maps from forms to contra-forms, contra-forms to forms, and a self-inverse map between spatial forms respectively,
\bea{
	*_\uparrow\lB\bm\mu^{(m)}\rB
	&= \star\lB \bm{{\bar V}} \wedge \bm\mu^{(m)} \rB^\sharp
	= \frac{1}{(d+1-m)!} \lb\frac{1}{m!} \mu_{\mu_1\ldots \mu_m} \ve_{\uparrow}^{\mu_1\ldots \mu_m \nu_1\ldots \nu_{d+1-m}} \rb
	\dow'_{\nu_1} \wedge \ldots \wedge \dow'_{\nu_{d+1-m}}, \nn\\
	*^\downarrow\lB\bm\mu^{[m]}\rB
	&= \star\lB\bV \wedge \bm\mu^{\flat(m)}\rB
	= \frac{1}{(d+1-m)!}  \lb\frac{1}{m!} \mu^{\mu_1\ldots \mu_m} \ve^{\downarrow}_{\mu_1\ldots \mu_m \nu_1\ldots \nu_{d+1-m}} \rb
	\df x^{\nu_{1}} \wedge \ldots \wedge \df x^{\nu_{d+1-m}},\nn\\
	*\lB\bm\mu^{(m)}\rB 
	&= \star \lB \bV \wedge \bm u \wedge \bm\mu^{(m)} \rB
	= \frac{1}{(d-m)!} 
	\lb\frac{1}{m!} \mu^{\mu_1\ldots \mu_m} \ve_{\mu_1\ldots \mu_m \nu_1\ldots \nu_{d-m}} \rb
	\df x^{\nu_{1}} \wedge \ldots \wedge \df x^{\nu_{d-m}}.
}
One can check that $**= - \sgn(G) (-)^{m(d-m)} $ and $*^\downarrow *_{\uparrow}= *_{\uparrow} *^\downarrow = - \sgn(G) (-)^{m(d+1-m)}$. 
Finally we need to define integration for NC full rank forms and contra-forms,
\bea{
	\int_{\cM_{(d+1)}} \bm\mu^{(d+1)} 
	&= \sgn(G)\int_{\cM_{(d+2)}} \bm{{\bar V}} \wedge \bm\mu^{(d+1)} 
	= \sgn(\g)\int \lbr \df x^\mu \rbr \sqrt{|\g|} \ *_\uparrow\lB\bm\mu^{(d+1)}\rB, \nn\\
	\int_{\cM_{(d+1)}} \bm\mu^{[d+1]} 
	&= \sgn(G)\int_{\cM_{(d+2)}} \bV \wedge \bm\mu^{\flat(d+1)} 
	= \sgn(\g)\int \lbr \df x^\mu \rbr \sqrt{|\g|} \ *^\downarrow\lB\bm\mu^{[d+1]}\rB,
}
where $\g_{\mu\nu} = p_{\mu\nu} +  n_\mu  n_\nu$ and $\g = \det \g_{\mu\nu} = - G$. Obviously a full rank spatial form would be zero. Rest of the notations and conventions follow from our relativistic discussion.

\subsection*{Non-covariant Differential Forms}

Choosing a non-covariant basis given in \cref{non_cov}, a vector and a one-form can be decomposed as,
\bea{
	\cX^M \dow_M &= 
	- \E{\F} \lb \cX_t +  B_t \cX_{\sss\sim} \rb \dow_{\sss\sim} 
	- \E{\F} \cX_{\sss\sim} \lb
		 B_t \dow_{\sss\sim} 
		+ \dow_t 
	\rb
	+  \cX^i \lb 
		\dow_i 
		- a_i \dow_t
		+ \lb B_i - a_i  B_t \rb \dow_{\sss\sim}
	\rb, \nn\\
	\cY_M \df x^M &= \cY_{\sss\sim} \lb \df x^{\sss\sim} -  B_\mu \df x^\mu \rb 
	+ \lb
		\cY_{\sss\sim}  B_t
		+ \cY_t 
	\rb \lb \df t + a_i \df x^i\rb
	+ g_{ij} \cY^j \df x^i.
}
It immediately follows that a spatial differential form ($\cY_t = \cY_{\sss\sim} = 0$) is indeed a pure differential form on the spatial slice. Such a form can be expanded in coordinate basis as,
\bee{
	\bm\mu^{(m)} = \frac{1}{m!}\mu_{i_1i_2\ldots i_m} \df x^{i_1} \wedge \df x^{i_2} \wedge \ldots \wedge \df x^{i_m}.
}
Since there exists a invertible metric $g_{ij}$ on this slice, there is a map between forms and contra-forms. One can check that the volume element $\bm\ve^{(d)}$ defined before is indeed a full rank form on the spatial slice and can be written in this setting as,
\bee{
	\bm \ve^{(d)} = \star[\bV \wedge \bm{{\bar V}}] = \frac{1}{d!} \lb V^{M} \bar V^N \e_{MN i_{1}\ldots i_{d}} \rb 
	\df x^{i_{1}} \wedge \ldots \wedge \df x^{i_{d}}.
}
The Hodge dual $\ast$ associated with it serves as Hodge dual operation on the spatial slice. Finally a full rank spatial form can be integrated on a spatial slice,
\bee{
	\int_{\cM_{(d)}} \bm\mu^{(d)} 
	= \sgn(G)\int_{\cM_{(d+2)}} \E{\F} \bm{{V}} \wedge \bm{{\bar V}} \wedge \bm\mu^{(d)} 
	= \sgn(g)\int \lbr \df x^\mu \rbr \sqrt{|g|} \ *[\bm\mu^{(d)}].
}
Here $g = \det g_{ij} = \E{2\F} \g = - \E{2\F} G$. Other conventions and notations are same as relativistic case.

\section{Comments on the Relativistic Entropy Current} \label{belinEC}

In this appendix we want to make some comments on the entropy current for a relativistic fluid. To
make more contact with \cite{Haehl:2015pja}, in this section we
consider the relativistic manifold $\cM_{(2n)}$ to be $2n$ dimensional, and denote indices on it by
$\mu,\nu\ldots$. On the local flat space $\bbR^{(2n-1,1)}$ however we denote indices by
$\a,\b\ldots$. This setup is equipped with a Vielbein $e^\a_{\ \mu}$, an affine connection
$\G^\l_{\ \mu\nu}$, a spin connection
$C^\a_{\ \mu\b} = e_{\b}^{\ \r} \lb \G^\nu_{\ \mu\r} e^\a_{\ \nu} - \dow_\mu e^a_{\ \r} \rb$, and a
non-abelian gauge field $A_\mu$. Correspondingly we have torsion tensor $\rmT^\a_{\ \mu\nu}$,
Riemann curvature tensor $R_{\mu\nu}{}^\a{}_\b$ and gauge field strength $F_{\mu\nu}$. $\mu,\nu\ldots$ can be
raised/lowered by metric $g_{\mu\nu}$, $\a,\b\ldots$ indices can be raised/lowered by flat metric
$\eta_{\a\b}$, while both type of indices can be interchanged by the Vielbein. The covariant
derivative on the other hand is given by $\N_\mu$ which is associated with all the connections. We
take the fluid data to be $\p_\b = \{ \b^\mu, [\L_{\Sigma(\b)}]^\mu_{\ \nu}, \L_{(\b)} \}$. In terms
of it we define fluid temperature $T = (-\b^\mu\b_\mu)^{-1/2}$, fluid velocity $u^\mu = T \b^\mu$,
scaled chemical potential $\nu = \L_\b + \b^\mu A_\mu$, chemical potential $\mu = T\nu$, scaled spin
chemical potential $[\nu_{\sss\Sigma}]^\a_{\ \b} = [\L_{\Sigma}]^\a_{\ \b} + \b^\mu C^\a_{\ \mu\b}$,
and spin chemical potential $[\mu_{\sss\Sigma}]^\a_{\ \b} = T [\nu_{\sss\Sigma}]^\a_{\ \b}$. Finally
we have canonical EM tensor $T^\mu{}_{\a}$, spin current $\Sigma^{\mu\a}{}_\b$, charge current
$J^\mu$, entropy current $J^\mu_S$, Belinfante EM tensor $T^{\mu\nu}_{(b)}$, and Belinfante (usual)
entropy current $J^\mu_{S(b)}$.

We wrote an off-shell generalization for the second law of thermodynamics in Cartan formalism in
\cref{hydro}, which in aforementioned notation will become,
\bem{\label{offshell_2ndLaw_app}
	\underline{\N}_{\mu} J_s^{\mu} 
	+ \b^{\nu} \lb 
		\underline{\N}_{\mu} T^{\mu}{}_{\nu}
		- \rmT^{\a}_{\ \nu\mu} T^{\mu}{}_{\a} 
		- R_{\nu\mu}{}^{\a}{}_{\b} \Sigma^{\mu\b}{}_{\a}
		- F_{\nu\mu}\cdot J^{\mu}
	\rb \\
	+ [\nu_{\sss\Sigma}]_{\b\a} \lb \underline{\N}_{\mu} \Sigma^{\mu\a\b} - T^{[{\b\a}]} - \Sigma_{\rmH }^{\perp \a\b} \rb
	+ \nu\cdot \lb \underline{\N}_{\mu} J^{\mu} - \rmJ_{\rmH }^\perp \rb \geq 0,
}
where $\underline{\N}_\mu = \N_\mu - \rmT^\nu_{\ \nu\mu}$. $\Sigma_\rmH^{\perp \a\b}$,
$\rmJ_\rmH^\perp$ are the anomalous Hall currents, which are determined in terms of an anomaly
polynomial $\bcP_{\rmC\rmS}^{(2n+2)}$,
\bee{
  \star_{(2n+1)}\bm\Sigma_\rmH{}^{\a\b} = \frac{\dow \bcP^{(2n+2)}_{\text{CS}}}{\dow \bR_{\b\a}}, \qquad
  \star_{(2n+1)}\bm\rmJ_\rmH = \frac{\dow \bcP^{(2n+2)}_\text{CS}}{\dow \bF}.
}
On imposing equations of motion \cref{EOM_usual} (after appropriate change of notation) this will
boil down to the second law of thermodynamics $\underline{\N}_\mu J^\mu_S \geq 0$. To compare this statement with that of \cite{Loganayagam:2011mu} we making a field redefinition,
\bee{
	[\nu_{\sss\Sigma}]_{\mu\nu} \ra [\nu'_{\sss\Sigma}]_{\mu\nu} = [\nu_{\sss\Sigma}]_{\mu\nu} +
        e_{{\a}[\mu} \d_\b e^{\a}_{{\ \nu}]} = \N_{[\nu} \b_{{\mu}]} + \rmT_{[{\sss \mu\nu}]\sss \r} \b^{\sss \r},
}
where $\d_\b$ is the diffeo, spin and gauge transformation associated with $\p_\b$. This field
redefinition does not spoil our equilibrium frame as the perturbation vanishes on promoting $\p_\b$
to an isometry. Further setting torsion to zero, this statement boils down to the statement of \cite{Loganayagam:2011mu},
\bee{
	\N_{\sss \mu} J_{S(b)}^{\mu}
	+ \b_{\nu} \lb 
		{\N}_{\mu} T^{\mu\nu}_{(b)}
		- F^{\nu\mu}\cdot J_\nu
		- \N_{\mu} \Sigma_{\rmH }^{\perp \nu\mu}
	\rb
	+ \nu\cdot \lb {\N}_{\sss \mu} J^{\mu} - \rmJ_{\rmH }^\perp \rb \geq 0,
}
where we have defined the \emph{Belinfante EM tensor},
\bee{\label{BEM}
  T^{\mu\nu}_{(b)} = T^{(\mu\nu)} + 2\N_\r \Sigma^{(\mu\nu)\r},
}
and \emph{Belinfante entropy current}\footnote{The motive of calling $S_{(b)}^{\sss M}$ Belinfante entropy current is primarily to distinguish it from $S^{\sss M}$, and secondly to relate it more closely to Belinfante EM tensor $T^{\sss MN}_{(b)}$. We couldn't find any existing name in the literature for this quantity.},
\bee{\label{BEC}
	J_{S(b)}^{\mu} = J_S^{\mu} + \b_{\nu} \lb \N_{\r} \Sigma^{\r\mu\nu} + T^{[{\mu\nu}]} - \Sigma_{\rmH }^{\perp \mu\nu} \rb,
}
which is more natural quantity to use when working with Belinfante EM tensor $T^{\sss
  \mu\nu}_{(b)}$. Note that the two entropy currents differ only off-shell and  boil down to the same when
the spin equation of motion is imposed,
\bee{
  \N_{\r} \Sigma^{\r\mu\nu} = T^{[{\nu\mu}]} + \Sigma_{\rmH }^{\perp \mu\nu}.
}
For comparison with \cite{Haehl:2015pja} we would be interested in
spinless relativistic fluids. In absence of anomalies we could define spinless fluids by $\Sigma^{\r\mu\nu} =
T^{[\mu\nu]} = 0$, but anomalies would not allow us to make this simple choice. Nevertheless we can
define spinless fluids by $\Sigma^{\r\mu\nu}$, $T^{[\mu\nu]}$ being totally determined by anomalies.

The transgression form business does not change much in Vielbein formalism. The end result is that
we can define certain quantities in terms of anomaly polynomial $\bcP^{(2n+2)}_{\text{CS}}$ and hatted connections $\bm{{\hat
    A}} = \bA + \mu \bm u$, $\bm{{\hat C}}{}^\a_{\ \b} = \bC^\a_{\ \b} +
[\mu_{\sss\Sigma}]^\a_{\ \b} \bm u$,
\bee{\nn
  \star\bm\Sigma^{\a\b}_{\bcP} = \frac{\bm u}{\df \bm u} \wedge \lb \frac{\dow
    \bcP}{\dow \bR_{\b\a}}^{(2n+2)} - \frac{\dow \bm{{\hat\cP}}{}^{(2n+2)}}{\dow \bm{{\hat R}}_{\b\a}} \rb, \qquad
  \star\bJ_{\bcP} = \frac{\bm u}{\df \bm u} \wedge \lb \frac{\dow
    \bcP^{(2n+2)}}{\dow \bF} - \frac{\dow \bm{{\hat\cP}}{}^{(2n+2)}}{\dow \bm{{\hat F}}} \rb}
\bee{
  \star \bm q_{_{\bcP}} = - \frac{\bm u}{\df \bm u} \wedge \lB \bcP^{(2n+2)}_{\text{CS}} - \bm{{\hat
    \cP}}_{\text{CS}} + \df \bm u \wedge \lb [\mu_{\sss\Sigma}]^\a_{\ \b} \frac{\dow
  \bm{{\hat\cP}}{}^{(2n+2)}}{\dow \bm{{\hat R}}^\a_{\ \b}} + \mu\cdot \frac{\dow \bm{{\hat\cP}}{}^{(2n+2)}}{\dow \bm{{\hat F}}}  \rb \rB.
}
In terms of these, anomalous sector of constitutive relations is given as,
\bee{
  (T^{\mu\a})_\rmA = q^\mu_{_{\bcP}} u^\a + q^\a_{_{\bcP}} u^\mu, \qquad
  (\Sigma^{\mu\a\b})_\rmA = \Sigma_{\bcP}^{\mu\a\b}, \qquad
  (J^\mu)_\rmA = J^\mu_{\bcP}.
}
These currents follow the Bianchi identities\footnote{Upon using the definition of Belinfante EM
  tensor from \cref{BEM}, and setting torsion to zero, these Bianchi identities reproduce the ones
  given in \cite{Haehl:2015pja}.},
\bea{\label{anom_EOM_app}
	\underline{\N}_{\mu} (T^{\mu}{}_{\nu})_\rmA
	&= 
	\rmT^{\a}_{\ \nu\mu} (T^{\mu}{}_{\a})_{\rmA}
	+ R_{\nu\mu}{}^{\a}{}_{\b} (\Sigma^{\mu\b}{}_{\a})_\rmA
	+ F_{\nu\mu} \cdot (J^{\mu})_{\rmA}
	- u_{\nu} \lb
		\mu \cdot \hat \rmJ_{\rmH}^\perp
		+ [\mu_{\sss\Sigma}]_{\a\b} \hat \Sigma_{\rmH }^{\perp \b\a}
	\rb, \nn\\
        &\qquad + \frac{1}{\sqrt{-g}} \d_\b \lb \sqrt{-g} T q_{_{\bcP}\nu} \rb \nn\\
	\underline{\N}_{\mu} (\Sigma^{\mu\a\b})_{\rmA} &= 
	\Sigma_{\rmH }^{\perp \a\b} - \hat \Sigma_{\rmH }^{\perp \a\b}, \nn\\
	\underline{\N}_{\mu} (J^{\mu})_{\rmA} &= \rmJ_{\rmH }^\perp - \hat \rmJ_{\rmH }^\perp.
}
Plugging these constitutive relations into the off-shell adiabaticity equation we can get a
relation for the entropy current,
\bee{
	\underline{\N}_{\mu} (J_S^{\mu})_\rmA \geq 0.
}
Hence the off-shell second law can be satisfied with a trivially zero entropy current 
\bee{
  J_S^\mu = 0.
}
In other words, entropy current $J_S^\mu$ does not get any contribution from anomalies. On the other hand,
using Bianchi identities in \cref{BEC}, we can read out the anomalous Belinfante entropy current,
\bee{
  (J^\mu_{S(b)})_\rmA = \b_\nu \hat\Sigma^{\perp \nu\mu}_\rmH,
}
which is what was found by \cite{Haehl:2015pja}. Note that $ \hat\Sigma^{\perp \nu\mu}_\rmH$ is
by definition antisymmetric in its last two indices, and differ from \cite{Haehl:2015pja} by a
factor of 2. Hence we have established that entropy current in Vielbein formalism $J_S^\mu$ does not
get contribution from anomalies, while the Belinfante entropy current does. Recall that a similar
situation appears for EM tensor as well; while the canonical EM tensor $T^{\mu\a}$ that appears in Vielbein
formalism is Noether current of translations, the symmetric Belinfante EM tensor $T^{\mu\nu}_{(b)}$
that appears in usual formalism
couples to metric in general relativity but does not correspond to any Noether current. Hence from
the point of view of symmetries, canonical EM tensor is a more natural quantity. On the same lines we
guess that Vielbein entropy current will be more naturally associated with the fundamental
$\rmU(1)_\rmT$ symmetry introduced by \cite{Haehl:2015pja}, as opposed to the Belinfante
entropy current. The former being independent of anomalies seems to strengthen this natural
guess. However one will have to do the explicit computation of $\rmU(1)_\rmT$ transformations
in presence of torsion to give any weight to this claim.

\bibliographystyle{utcaps.bst}
\bibliography{aj-bib}

\end{document}